\documentclass[a4paper,fleqn,usenatbib]{mnras}

\usepackage{newtxtext}

\usepackage[T1]{fontenc}
\usepackage{ae,aecompl}

\usepackage{graphicx}
\usepackage{amsmath}	
\usepackage{amssymb}	

\usepackage{color}
\title[A new outburst of $\rho$\,Cas]{A new outburst of the yellow hypergiant star $\rho$\,Cas}

\author[M. Kraus et al.]{M.~Kraus$^{1}$\thanks{E-mail: michaela.kraus@asu.cas.cz},
        I.~Kolka$^{2}$, A.~Aret$^{1,2}$, D.~H.~Nickeler$^{1}$, G.~Maravelias$^{3,4}$,
         T.~Eenm\"{a}e$^{2}$, 
        \newauthor A.~Lobel$^{5}$, V.~G.~Klochkova$^{6}$
            \\
\\
$^{1}$ Astronomick\'y \'ustav, Akademie v\v{e}d \v{C}esk\'e republiky,
Fri\v{c}ova 298, 251\,65 Ond\v{r}ejov, Czech Republic\\
$^{2}$Tartu Observatory, University of Tartu, 61602 T{\~o}ravere, Tartumaa, Estonia\\
$^{3}$ IESL, Foundation for Research and Technology-Hellas, 100 Nikolaou Plastira Street, 71110 Heraklion, Crete, Greece\\
$^{4}$ Physics Department, University of Crete, P.O. Box 2208, 71003 Heraklion, Crete, Greece\\
$^{5}$ Royal Observatory of Belgium, Ringlaan 3, 1180, Brussels, Belgium\\
$^{6}$ Special Astrophysical Observatory of the Russian Academy of Sciences, Nizhnii Arkhyz 369167, Russia
}

\date{Accepted XXX. Received YYY; in original form ZZZ}

\pubyear{2018}

\begin{document}
\label{firstpage}
\pagerange{\pageref{firstpage}--\pageref{lastpage}}
\maketitle

\begin{abstract}
Yellow hypergiants are evolved massive stars that were suggested to be in post-red 
supergiant stage. Post-red supergiants that evolve back to the blue, hot side of the 
Hertzsprung-Russell diagram can intersect a temperature domain in which their atmospheres 
become unstable against pulsations (the Yellow Void or Yellow Wall), and the stars can experience outbursts with short, but 
violent mass eruptions. The yellow hypergiant $\rho$\,Cas is famous for its historical and recent 
outbursts, during which the star develops a cool, optically thick wind with a very brief but high 
mass-loss rate, causing a sudden drop in the light curve. Here we report on a new outburst of $\rho$\,Cas 
which occurred in 2013, accompanied by a temperature decrease of $\sim$3000 K and a brightness drop of 
0.6\,mag. During the outburst TiO bands appear, together with many low excitation metallic atmospheric 
lines characteristic for a later spectral type. With this new outburst, it appears that the time 
interval between individual events decreases, which might indicate that $\rho$\,Cas is preparing for a 
major eruption that could help the star to pass through the Yellow Void.
We also analysed the emission features that appear during phases of maximum brightness and find that 
they vary synchronous with the emission in the prominent [Ca\,{\sc ii}] lines. We conclude that the
occasionally detected emission in the spectra of $\rho$\,Cas, as well as certain asymmetries seen in 
the absorption lines of low to medium-excitation potential, are circumstellar in nature, and we discuss 
the possible origin of this material.
\end{abstract}




\section{Introduction} \label{sec:intro}

Yellow Hypergiants (YHGs) are cool ($T_{\rm eff}$ = 4000--7000\,K), luminous (5.4 $\leq \log L/$L$_\odot 
\leq$ 5.8) massive stars. \citet{1998A&ARv...8..145D} proposed that these objects evolve
to the blue, hot side of the Hertzsprung--Russell (HR) diagram after 
having passed through the red supergiant (RSG) phase \citep{1998A&ARv...8..145D}. Despite the fact that 
they are amongst the visually brightest objects, YHGs are rare, indicating that this phase is short. 
Nevertheless, these objects are cornerstones in the evolution of massive stars, because they link the 
cool RSGs and the hot pre-supernova stages such as Luminous Blue Variables and Wolf-Rayet stars.

When the stars enter the temperature regime between 7000 and 11000\,K, which is referred to as the 
`Yellow Evolutionary Void' \citep{1998A&ARv...8..145D} or `Yellow Wall' \citep{2009ASPC..412...17O}, 
their atmospheres start to become dynamically unstable and strong mass loss sets in 
\citep{1995A&A...302..811N, 1997MNRAS.290L..50D, 2001ApJ...560..934S}. Moreover, when approaching this instability 
domain the stars may shed their outer layers in a series of outbursts denoted as 
`bouncing against the Yellow Void' \citep{1997MNRAS.290L..50D}. These outbursts can result in the formation of multiple 
detached shells such as the double-shell structure detected on mid-infrared images of the post-RSG star 
IRAS\,17163-3907 \citep[the `Fried Egg' nebula,][]{2011A&A...534L..10L}, or the ejecta resolved on HST 
images around the YHG star IRC\,+10420 \citep{2010AJ....140..339T}. During an outburst event, the star 
appears dimmer, hence cooler so that the object evolves on an apparent red loop in the HR diagram. 

The proposed bouncing might continue until a significant part of the outer 
shell is lost and the stellar atmosphere finally becomes stable again and the star might
appear on the blue side of the Yellow Void where it could continue its life as a blue supergiant of $\alpha$ Cygni-type 
variable or as a Luminous Blue Variable. Alternatively, if its mass-loss behaviour has changed from 
spherically to axis-symmetrically, as it seems to be the case for IRC\,+10420 
\citep*{2007ApJ...671.2059D} and possibly also for V509\,Cas \citep{2017ASPC..510..162A}, the star might 
also appear as B[e] supergiant.
  
YHGs hence occupy a critical phase in the evolution of massive stars, and the mass loss during this 
phase is crucial for the fate of the object. For a better comprehension of the physics that control the 
atmospheric dynamics and eventually lead to mass eruptions in YHGs, as well as to catch phases of
mass ejections, monitoring these stars is essential. Here, we report on the discovery of a new outburst 
in $\rho$\,Cas, which is one of the northern Galactic YHGs that we monitor since 2010. 

\begin{table}
\centering
\caption{Observation log of the Ond\v{r}ejov spectra}
\label{tab:obslog}
\begin{tabular}{lcrrr}
\hline 
Date & JD & T$_{\rm Exp}$ & T$_{\rm Exp}$ & T$_{\rm Exp}$ \\
 &  & (H$\alpha$) & ([Ca\,{\sc ii}]) & (Ca\,{\sc ii} trip.) \\
& (d) & (s) & (s) & (s)\\
\hline 
2010-07-17&  2455394&  600  & ---  & 1800  \\
2011-03-06&  2455627&  900  & 1100 & ---   \\   
2011-04-18&  2455669&  300  & 300  & ---   \\  
2011-08-21&  2455795&  900  & 800  & ---   \\
2011-08-24&  2455797&   --- & ---  &  600 \\
2011-09-04&  2455809&  600  & 1200 &  900  \\ 
2011-09-06&  2455811&  ---  & 900  & ---   \\
2011-09-11&  2455816&  1300 & 800  & ---   \\  
2011-09-12&  2455816&  ---  & ---  & 1580  \\
2011-09-15&  2455819&  900  & 1000 & 1200  \\ 
2011-09-23&  2455828&  ---  & 1800 & ---   \\
2011-09-24&  2455828&  1200 & ---  & ---    \\ 
2011-10-15&  2455850&  1800 & 1800 & 1800   \\
2011-11-06&  2455872&  900  & 1200 & 1800   \\
2012-07-28&  2456136&  200  &  300 & 1200   \\
2012-08-03&  2456143&  800  &  600 & 1000   \\
2012-09-28&  2456199&  900  &  800 & 1174   \\
2013-06-27&  2456471&  1200 & ---  & ---    \\
2013-06-28&  2456471&  ---  & 1200  & --- \\
2013-10-01&  2456567&  1200 &  1800 & 1800  \\
2014-07-19a&  2456857&  300  & ---  & ---    \\
2014-07-19b&  2456858&  300  & 750  & 1200   \\
2014-10-28&  2456959&  2400 & 3000  & 6000 \\
2014-12-13&  2457005&  900  & ---   & --- \\
2015-01-13&  2457036&  900  & 2600  & --- \\
2015-02-06&  2457060&  3600 & 3600  & --- \\
2015-03-19&  2457100&  600  & 900   & 1200 \\
2015-06-04&  2457178&  600  & ---   & --- \\
2015-06-05&  2457178&  ---  & 900   & 900 \\  
2015-06-12&  2457186&  1100 & ---   & --- \\
2015-06-13&  2457186&  ---  & 700   &  700  \\
2015-06-17&  2457191&  1500 &  ---  & --- \\
2015-06-26&  2457199&  600  & 600   & 600 \\
2015-07-09&  2457213&  1200 & 900   & 1200 \\
2015-07-20&  2457224&  400  & 400   & 400 \\
2015-11-24a&  2457350&  1200 &  --- & --- \\
2015-11-24b&  2457351&  800  & 1200 & 1300 \\
\hline 
\end{tabular}
\end{table}

\section{The Object}   
The YHG star $\rho$\,Cas (HD 224014) is famous for its historical and recent 
outbursts, and at least three have been recorded so far. The first major outburst occurred in 
1945--1947 and was discovered by Popper\footnote{D. M. Popper, Harvard Announcement Card, No. 752, 1946; 
for the light curve see \citet{1961ApJS....5..381B}.} (1946), who reported on the deep minimum in 1946. 
Another one took place in 2000--2001 \citep{2003ApJ...583..923L} and was henceforth referred to as the 
Millennium outburst. Each of these major outbursts was accompanied by a drop in the light curve by more
than one magnitude. A third, less violent incident was noted in 1985--1986 \citep*{1988TarOT..92...40B}, 
during which the brightness dimmed by only about 0.6--0.7\,mag. During every event TiO absorption bands 
arose in the optical and near-infrared stellar spectra \citep{1947AJ.....52..129P, 1948MNRAS.108..279T, 
1988TarOT..92...40B, 2003ApJ...583..923L}, indicative for the development of a cool, optically thick 
wind with a high mass-loss rate. 

Dust with a temperature of about 600\,K was detected with IRAS in 1983 consistent with an optically
thin dust shell formed from the ejected material during the 1946 outburst \citep{1990ApJ...351..583J}.
Recent mid-infrared photometric observations by \citet{2016AJ....151...51S} confirm the expansion and 
thinning of this dust shell, while no indication for new dust formation following the millenium 
outburst was found yet. Also, no evidence for extended emission resulting from previous mass-loss 
episodes during the RSG phase was found in deep {\it Hubble Space Telescope} images 
\citep*{2006AJ....131..603S}. 

During the quiescence phases, $\rho$\,Cas shows line-profile and low-amplitude photometric variability
that is ascribed to semi-regular pulsations \citep{1986PASP...98..914S} with a quasi-period of 300\,d 
and a corresponding brightness fluctuation of 0.2\,mag \citep{1991A&A...246..441Z}. These pulsations 
manifest themselves also in an excess line broadening of the photospheric lines. Pure radial pulsations 
as cause of the temperature and radial velocity variation have been ruled out 
\citep{1994A&A...291..226L}, and \citet{1998A&ARv...8..145D} proposed that $\rho$\,Cas pulsates in a 
combination of both pressure and gravity modes. 

Over the pulsation cycles, $\rho$\,Cas displays effective temperature oscillations between 500\,K 
\citep*{1999ApJ...523L.145I} and 750\,K \citep{1998A&A...330..659L}, although slightly larger 
values have recently been reported as well \citep{2014ARep...58..101K}. Moreover, the temperature 
follows very tightly the variations of the light curve, also during the outbursts 
\citep{2003ApJ...583..923L}.

Occasionally, numerous and prominent emission lines arise in the optical and infrared spectral regions. 
These appearances were found to be synchronized with phases of maximum light corresponding to periods of 
fast expansion of the atmosphere \citep[see, e.g.,][]{2003ApJ...583..923L, 2006ApJ...651.1130G, 
2007PASJ...59..973Y}. The molecular features of the CO first-overtone bands arising in the near-infrared 
longward of 2.3\,$\mu$m seem to be most sensitive. These bands change from intense emission to strong 
absorption on a time-scale of weeks to months. An explanation for their occurrence is, however, 
controversial: \citet{2006ApJ...651.1130G} place the CO band formation region in the vicinity of the 
pulsating photosphere and correlate their variability with the pulsation cycles, whereas 
\citet{2007PASJ...59..973Y} propose the CO band formation to be related to a mass-loss event in which 
the change from emission to absorption is due to the expansion of the ejected gas shell. Some emission 
lines in the spectrum of $\rho$\,Cas seem to be always present, such as the [Ca{\sc ii}] lines, 
\citep[e.g.,][]{2003ApJ...583..923L, 2017ASPC..508..239A}, which are proposed to form in a circumstellar 
gas shell.

A further peculiarity in the spectra of $\rho$\,Cas is the appearance of split lines. The origin of these 
lines is still controversial. \citet{1998A&A...330..659L} identify two types of split lines: in one type 
a static emission of the same transition superimposes the broad, cyclically variable absorption 
component. This splitting is observed in the cores of the absorption components of neutral and ionized 
metallic lines. These lines have typically (very) low excitation energies, and the static emission is 
proposed to form in an outer envelope \citep{1994A&A...291..226L} or tenuous, diffuse circumstellar gas 
shell \citep{1998A&A...330..659L}. A second type of split lines arises from the superposition of an 
emission line on the highly variable absorption profile of a different transition, causing an apparent
split of the absorption line. A superimposed circumstellar emission component was also the preferred 
explanation for the split lines seen by \citet{2006ApJ...651.1130G}. 

However, alternative scenarios were suggested as well. For instance, \citet{1992A&A...254..280G} 
interpreted the split lines as due to two superimposed absorption components of which the red one 
originates within the photosphere of $\rho$\,Cas, whereas the blue
component forms in a hot, expanding cicumstellar envelope. Support for such a scenario was found by 
\citet{2014ARep...58..101K} and \citet{2018ARep...62..623K} based on a comparison of the dynamics from symmetric, non-distorted 
absorption lines and from permanently split lines. 
Although the star has been subject 
to many studies and detailed analyses throughout the past century, its atmospheric dynamics causing unusually 
broad photospheric absorption lines, peculiar and strongly variable emission line spectra, recurring massive 
outburst events, pulsation properties, and characteristics of its circumstellar environment still remain 
elusive.

\section{Observations} \label{sec:obs}

We spectroscopically monitored $\rho$\,Cas between 2010 July 16 and 2015
November 24. The observations were obtained using the Coud\'{e} spectrograph
\citep{2002PAICz..90....1S} attached to the Perek 2-m telescope at Ond\v{r}ejov
Observatory. Until the end of May 2013, the observations were taken with the
830.77 lines\,mm$^{-1}$ grating and a SITe $2030\times 800$ CCD. Beginning in
June 2013, we used the newly installed PyLoN $2048\times 512$BX CCD. The
spectra were taken in three different wavelength regions: 6250--6760\,\AA,
6990--7500\,\AA, and 8470--8980\,\AA, and the spectral resolution in these
ranges are (with both CCDs) $R\simeq$ 13\,000, 15\,000, and 18\,000, 
respectively. These regions were chosen to cover several specific emission 
features expected from the environments of YHGs: H$\alpha$, [O\,{\sc i}] 
$\lambda\lambda$ 6300, 6364\,\AA, [Ca\,{\sc ii}] $\lambda\lambda$ 7291, 
7324\,\AA, and the Ca\,{\sc ii} $\lambda\lambda$ 8498, 8542, 8662\,\AA \ 
infrared triplet. Moreover, these regions cover many Fe\,{\sc i} and 
Fe\,{\sc ii} lines with different excitation potentials, the hydrogen Paschen 
lines Pa(12)--Pa(18), as well as numerous other metal lines of different 
excitation and ionization states suitable for detailed studies of the
atmospheric motions of the star. The log of the observations is given in 
Table\,\ref{tab:obslog}.

For wavelength calibration, a comparison spectrum of a ThAr lamp was taken
immediately after each exposure. The stability of the wavelength scale was
verified by measuring the wavelength centroids of [O\,{\sc i}] sky lines. The
velocity scale remained stable within 1\,km\,s$^{-1}$.

\begin{table}
\centering
\caption{Observation log of the echelle spectra}
\label{tab:echelle}
\begin{tabular}{cccc}
\hline 
Date & JD & Range & Inst. \\
& (d) & (\AA) & \\
\hline 
2010-09-24 & 2455464 & 5216--6690 & NES \\
2011-01-13 & 2455574 & 5208--6683 & NES \\
2011-09-14 & 2455819 & 3985--6980 & NES \\
2012-11-22 & 2456254 & 3770--9000 & HERMES \\
2013-02-02 & 2456326 & 3916--6980 & NES \\
2013-10-31 & 2456597 & 3770--9000 & HERMES \\
2014-10-01 & 2456931 & 5417--8479 & NES \\
2014-10-04 & 2456935 & 3770--9000 & HERMES \\
\hline
\end{tabular}
\end{table}

All data were reduced and heliocentric velocity corrected using standard
IRAF\footnote{IRAF is distributed by the National Optical Astronomy
Observatories, which are operated by the Association of Universities for
Research in Astronomy, Inc., under cooperative agreement with the National
Science Foundation.} tasks. We also observed once per night a rapidly rotating
star (HR\,7880, Regulus, $\zeta$\,Aql) as a telluric standard to perform the
telluric correction.

Our data are supplemented by five spectra obtained with the
high-resolution ($R\sim 60\,000$) Nasmyth Echelle Spectrograph \citep[NES, 
see][]{2017ARep...61..820P} attached to the 6-m telescope at the Special 
Astrophysical Observatory (SAO) in Russia. The observations spread from 2010 
September 24 to 2014 October 4. Details on the spectrograph and on the reduction 
procedure are provided by \citet{2014ARep...58..101K}.

Three high-resolution spectra \citep[R=85\,000;][]{2011A&A...526A..69R} were obtained
using the High Efficiency and Resolution Mercator Echelle Spectrograph
(HERMES) mounted on the Mercator 1.2-m telescope at the Roche de los
Muchachos Observatory, La Palma, Spain. The HERMES spectra were observed on
2012 November 22, 2013 October 31, and 2014 October 4. They have been calibrated with
latest version of the HERMES pipeline for the typical echelle calibration
steps of wavelength calibration, order-extraction, flat-fielding, background
subtraction, and order merging. The quality of the echelle dispersion
solutions was tested with the position of sharp telluric lines to check the
wavelength scales are accurate. The spectra of October 2013 and 2014 consist of
two subsequent exposures that were co-added for increasing the
signal-to-noise ratios. The resulting flux spectra were normalized to the
continuum flux level around selected spectral lines and in some spectral
regions of interest.

\begin{figure*}
\begin{center}
\includegraphics[width=\hsize,angle=0]{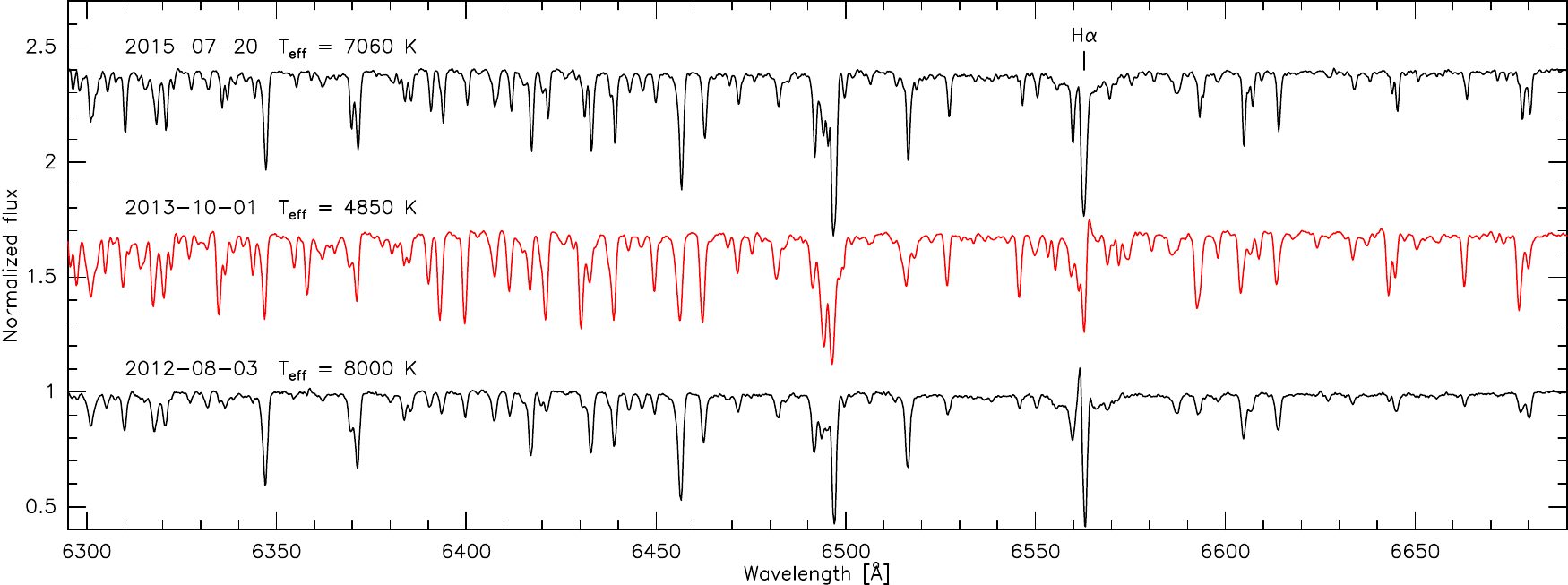}
\includegraphics[width=\hsize,angle=0]{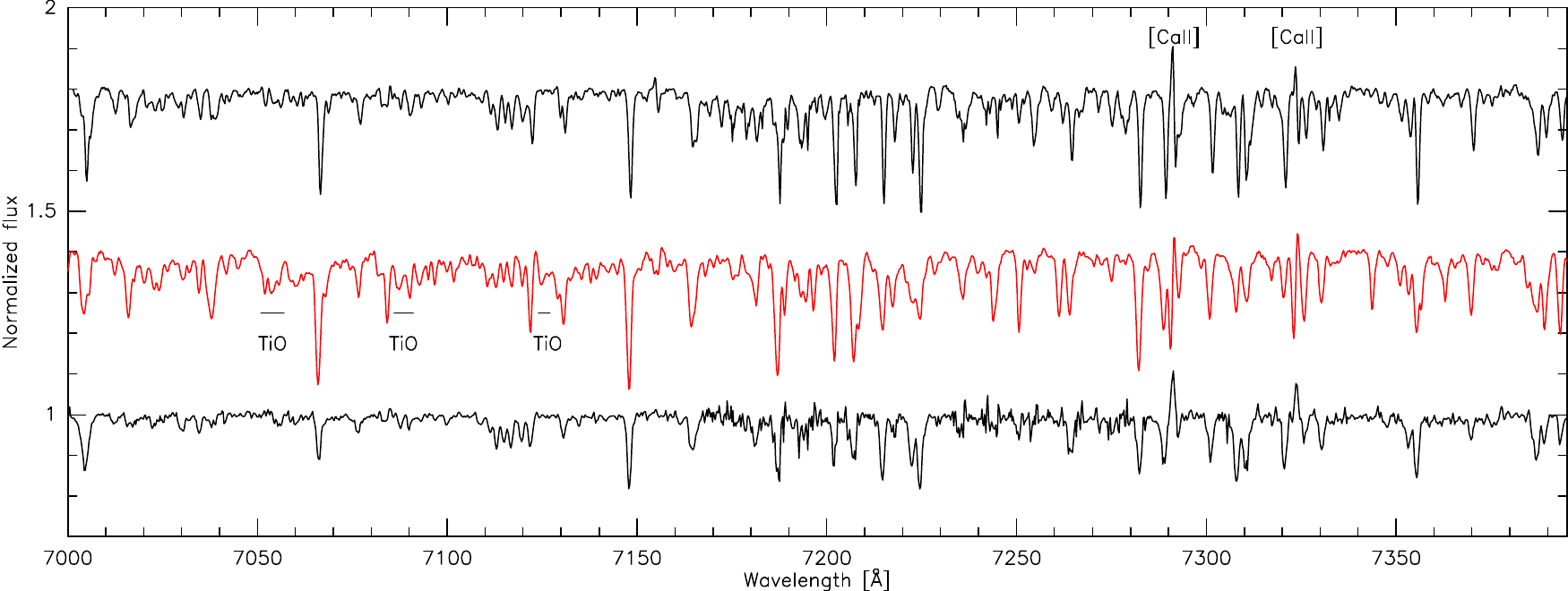}
\includegraphics[width=\hsize,angle=0]{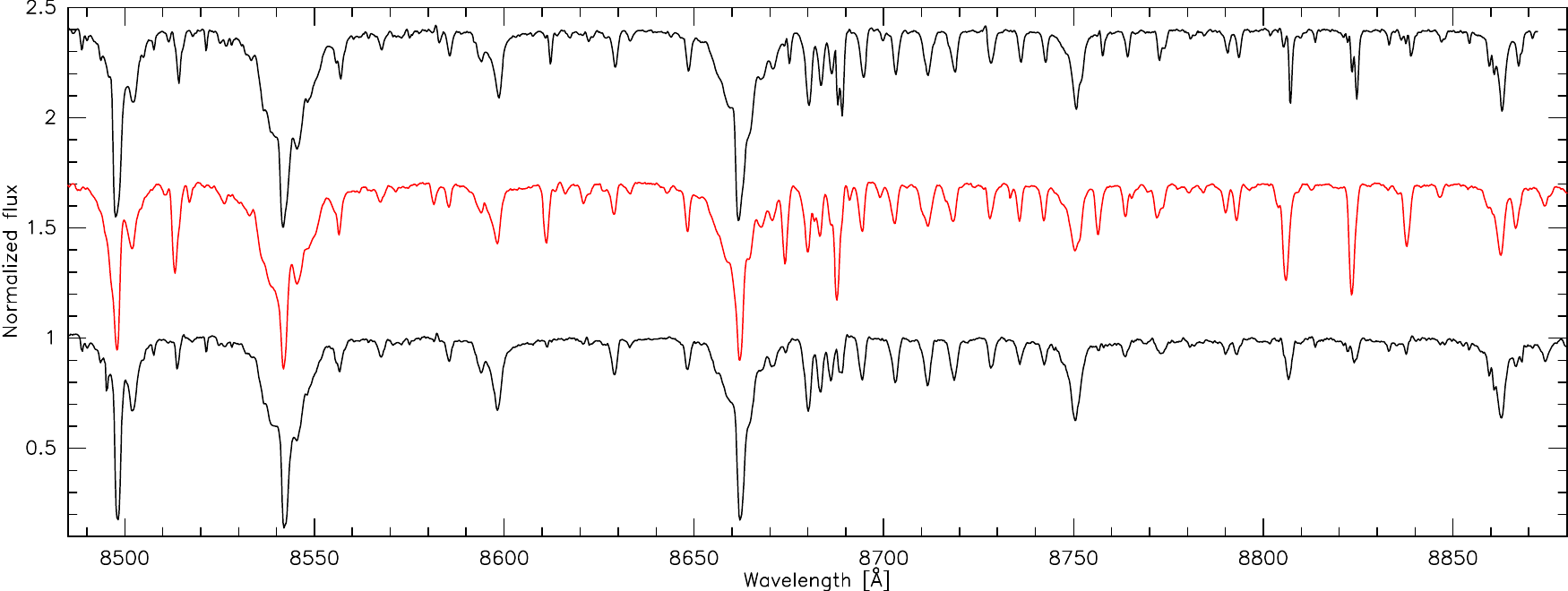}
\caption{Ond\v{r}ejov spectra during the hottest (bottom spectrum), outburst (coolest, middle spectrum) 
and recovered (top spectrum) phases in three different spectral regions. Prominent emission is seen in 
H$\alpha$ and in [Ca\,{\sc ii}] $\lambda\lambda$7291,7324. During outburst, weak TiO absorption appears. 
The positions and widths of the band heads are indicated. For best visualization the spectra in each 
wavelength region are offset arbitrarily.}
\label{fig:spec} 
\end{center}
\end{figure*}

\begin{figure*}
\begin{center}
\includegraphics[width=0.49\hsize,angle=0]{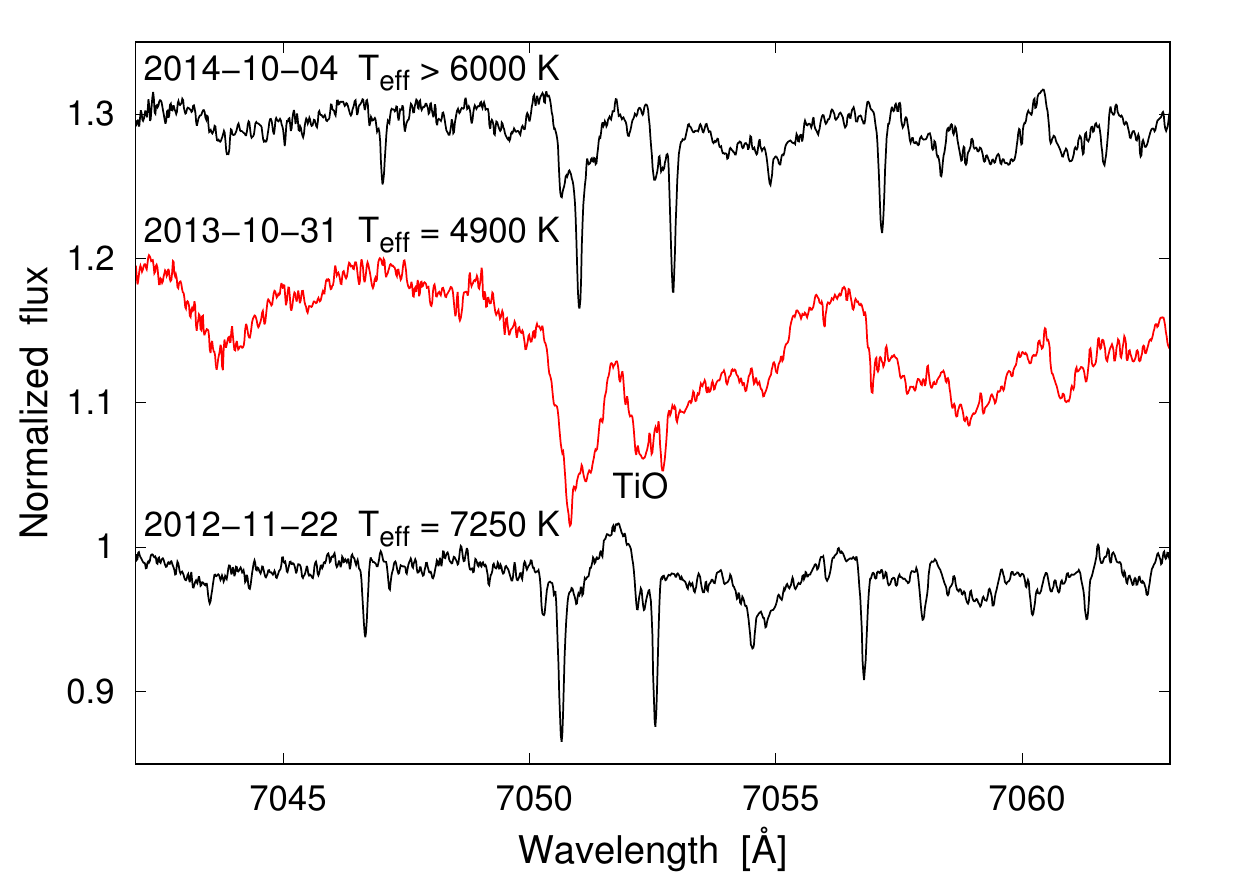}
\includegraphics[width=0.49\hsize,angle=0]{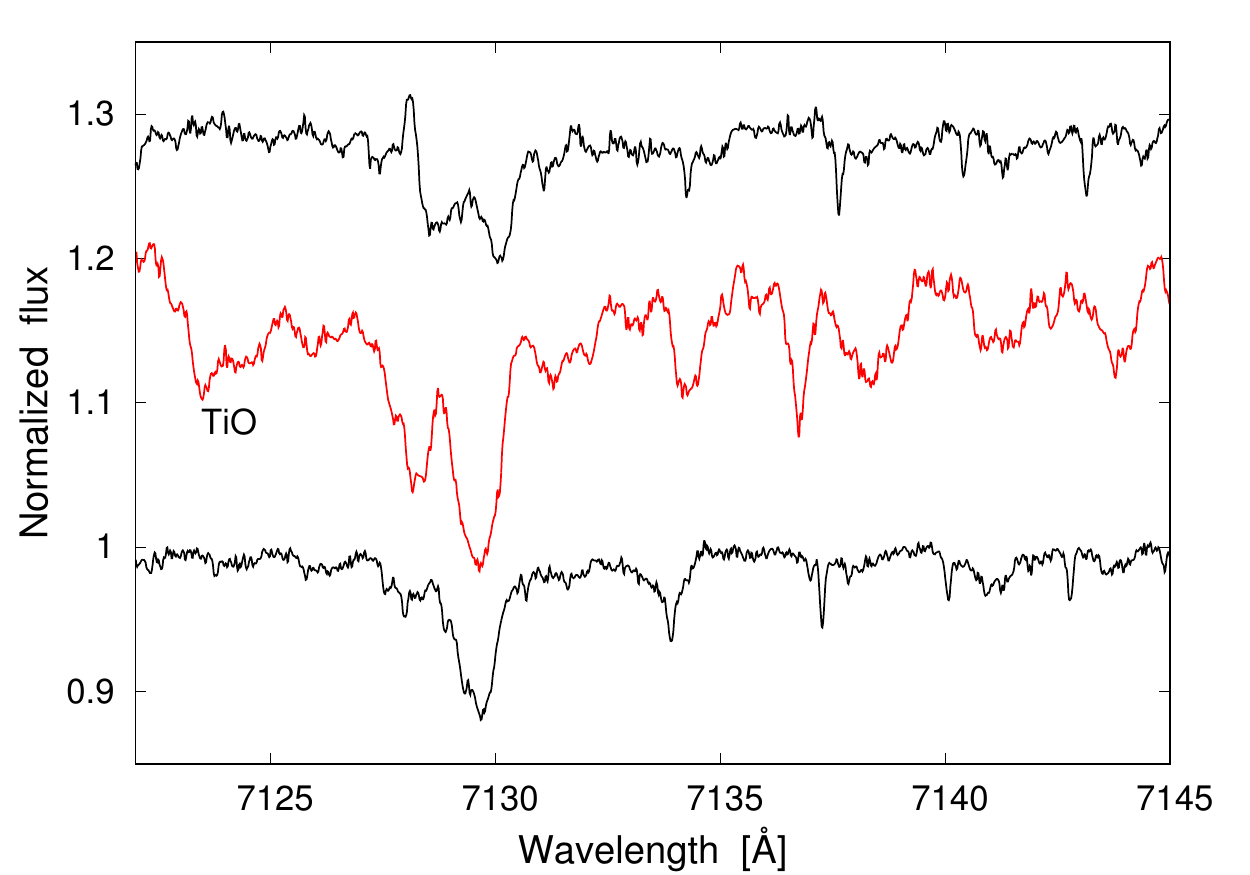}
\caption{Clear presence of TiO bands in the HERMES data during the outburst in October 2013 compared 
to observations taken prior and later.}
\label{fig:TiO-HERMES}
\end{center}
\end{figure*}

The log of the echelle observations is given in Table\,\ref{tab:echelle}.

All spectra were finally corrected for the systemic velocity of $\rho$\,Cas and normalized to the 
continuum. For the systemic velocity we adopted a value of $-47$\,km\,s$^{-1}$, as was previously found 
by \citet{2003ApJ...583..923L}, \citet{2014ARep...58..101K}, and \citet{2018ARep...62..623K}.

\section{Results} \label{sec:results}

Our data collection spreads over about 4.5 years. The most prominent emission features in the spectra 
are H$\alpha$, in which we observe gradual changes in its blue and/or red emission wings, and the 
persistent, though weak emission in the [Ca{\sc ii}] $\lambda\lambda$7291,7324\,\AA \ lines. None of our 
spectra displays indication for [O{\sc i}] $\lambda\lambda$6300\,\AA \ emission, in contrast to the 
hotter YHG counterparts IRC\,+10\,420 or V509\,Cas in which prominent emission in both sets of forbidden  
lines is usually seen \citep[e.g.,][]{2017ASPC..508..239A}.

During the first half of our monitoring period we observe long-term line-profile variability in both the 
radial velocity and the shape of all lines. Such spectral variabilities for the atmospheric dynamics of 
$\rho$\,Cas in quiescence have been reported by many investigators \citep[see][for comprehensive 
overviews]{2003ApJ...583..923L, 2006ApJ...651.1130G, 2014ARep...58..101K}  and generally comply 
with the interpretation of a combination of slow radial and non-radial pulsations on time-scales of 
several hundred days \citep[e.g.,][]{1934AN....253..457H, 1986PASP...98..914S, 1991A&A...246..441Z, 
1998A&A...330..659L, 2000PASP..112..363P}. 

In the following, we focus mainly on the second half of our observing epoch which is dominated by the
new outburst. We present the various indications for the outburst and describe the post-outburst phase. 
We discuss the observed spectral features in connection and comparison with the reported dynamical 
characteristics of $\rho$\,Cas in the past. For illustration purposes, we present the full 
time-evolution of several selected lines in Figs.\,\ref{fig:app1}--\ref{fig:app4} of the Appendix.

\subsection{New outburst in 2013}

In 2013 the spectroscopic behaviour of $\rho$\,Cas changed noticeably. In June, the cores of the lines 
of neutral elements such as Fe{\sc i} and Ti{\sc i} as well as the low- to intermediate-excitation lines 
of singly ionized elements appear strongly blue-shifted and additionally display a high-velocity 
blue-shifted wing (Figs.\,\ref{fig:app1} and \ref{fig:app2}). In October, all these lines display deep 
blue-shifted absorption, while at the same time the lines of medium and high-excitation of ionized 
elements like Fe{\sc ii} and Si{\sc ii} weaken and many previously not seen low excitation lines of 
neutral metals appear. This significant change in spectral appearance is demonstrated in Figure 
\ref{fig:spec} where we show three spectra taken in the years 2012, 2013, and 2015 in the three covered 
wavelength regions. This drastic change in the strengths of the lines and the appearance of many 
additional low excitation lines of neutral metals indicate that during fall 2013 the atmosphere of 
$\rho$\,Cas resembled that of a considerably later spectral type. 

Support for such an interpretation comes from the appearance of TiO absorption bands in the red 
spectral region in October 2013. This can be seen from inspection of the spectra 
shown in the middle panel of Fig.\,\ref{fig:spec}. For better visualization, the positions 
and widths of the TiO band heads are indicated by horizontal bars. These bands are also 
seen in the high-resolution spectra taken 30 days later with HERMES (Figure \ref{fig:TiO-HERMES}), 
confirming their identification in the lower resolution Ond\v{r}ejov data. No TiO bands were seen 
in the spectra taken before or after.

The appearance of TiO bands together with the changes in the strengths and profile shapes of the 
photospheric lines listed above are clear signs for a (much) cooler atmosphere during this epoch. So 
far, these spectral characteristics were seen in $\rho$\,Cas only during its outbursts in the years 
1946-47 \citep{1947AJ.....52..129P, 1948MNRAS.108..279T}, 1986 \citep{1988TarOT..92...40B}, and the 
millenium outburst 2000--01 \citep{2003ApJ...583..923L}, indicating that during 2013 $\rho$\,Cas 
underwent another outburst. 

\begin{figure*}
\begin{center}
\includegraphics[width=0.9\hsize,angle=0]{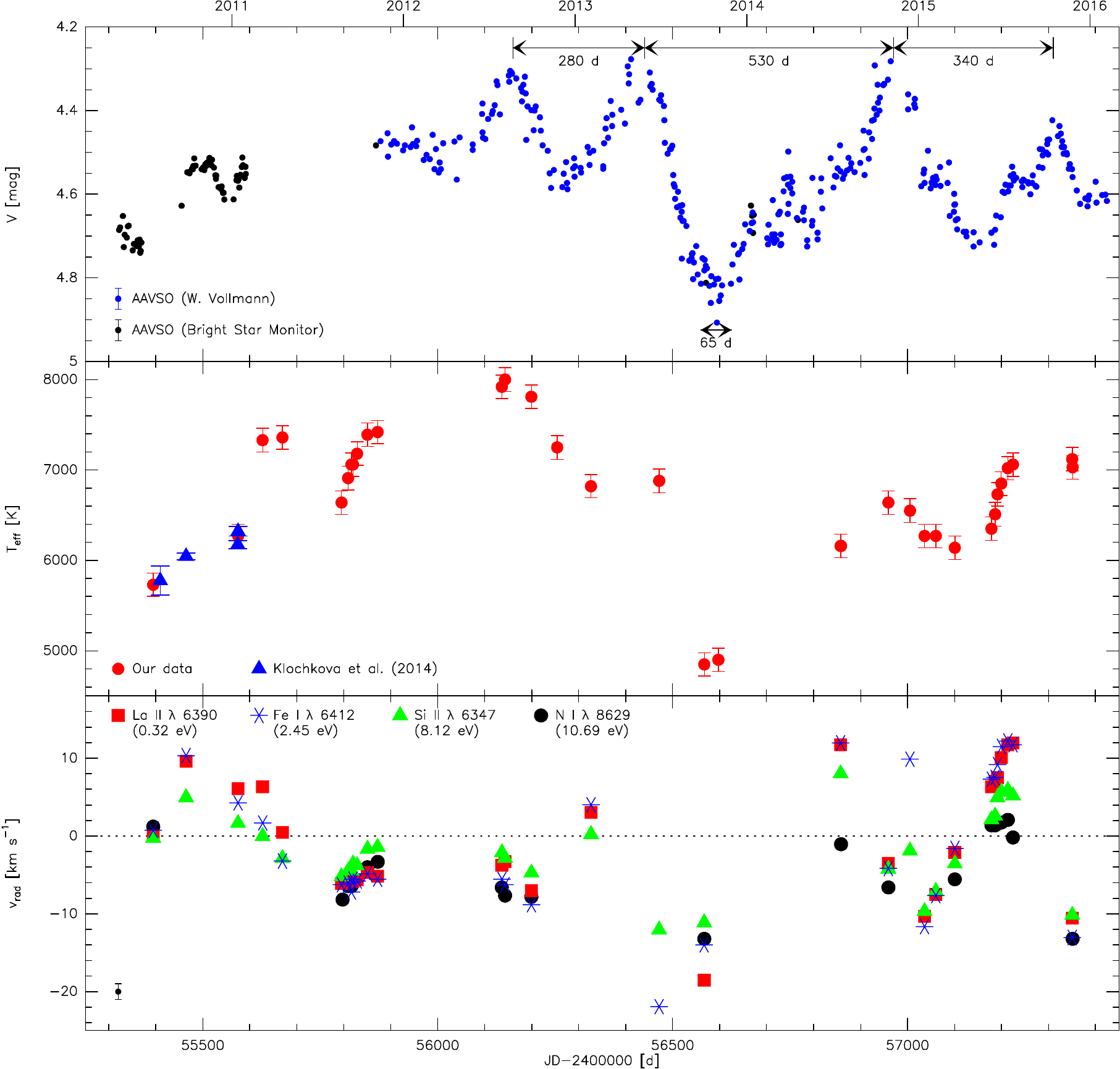}
\caption{Top: $V$-band light curve of $\rho$\,Cas.  
Middle: Effective temperature variation derived from the Fe\,{\sc i}  $\lambda$6431/Fe\,{\sc ii}  
$\lambda$6433 line depth ratio. Bottom: Radial velocity variations in selected photospheric lines.
Typical errorbars are indicated in each panel.}
\label{fig:photom}
\end{center}
\end{figure*}

All previous outbursts of $\rho$\,Cas were connected with a significant drop in the light curve. To 
check the brightness behaviour during our spectroscopic observing campaign from July 2010 until the end 
of 2015, we constructed the light curve of $\rho$\,Cas based on $V$-band magnitudes retrieved from the 
AAVSO\footnote{\href{https://www.aavso.org/}{https://www.aavso.org/}} database. For a coherent set of 
data, we use only the light curve measurements obtained by Wolfgang 
Vollmann\footnote{\href{https://bav-astro.eu/rb/rb2015-1/23.pdf}{https://bav-astro.eu/rb/rb2015-1/23.pdf}}, and 
complement them with data from the AAVSO Bright Star Monitor database. All values have been 
transformed to standard Johnson $V$ magnitudes. To account for small differences in the used filters, 
we shifted the values from Vollmann by $+$0.1 mag to match them to the data of the Bright Star Monitor.
The combined $V$-band light curve is shown in the top panel of Figure \ref{fig:photom}. Typical 
errorbars range from 0.02 to 0.03 mag as indicated in the lower left corner.

The light curve displays three phases of maximum brightness of about 4.3 mag that were reached in 
August 2012, May 2013, and November 2014, and an additional one in October 2015, which was 
$\sim$0.15 mag dimmer. The time spans between consecutive maxima are with approximately 280\,d, 530\,d, 
and 340\,d rather irregular, however, they comply with former determinations of photometric periods of 
280--300\,d \citet{1991A&A...246..441Z} and \citet{2000PASP..112..363P}, and of a radial velocity
periodicity of 520\,d \citep[e.g.,][]{1986PASP...98..914S}. 

We also note the sudden drop in brightness by 0.55 mag in the year 2013 between end of May, 
when the star was still at maximum brightness, and mid October, exactly at the time when we observe
the presence of TiO bands in our red spectra. This pronounced minimum lasted for about 65\,d. After 
that, the star started its recovery during which the brightness increased much shallower than it 
declined. The full recovery up to the next maximum took about a year. It is worth mentioning that 
the $V$ brightness minimum in fall 2013 did not decrease to below the 5th magnitude, which is unlike the three previous outbursts recorded for $\rho$\,Cas where the star became
much dimmer ($V>$ 5.0\,mag)

\subsection{Temperature variation} \label{subsec:teff}

\begin{figure}
\begin{center}
\includegraphics[width=\hsize,angle=0]{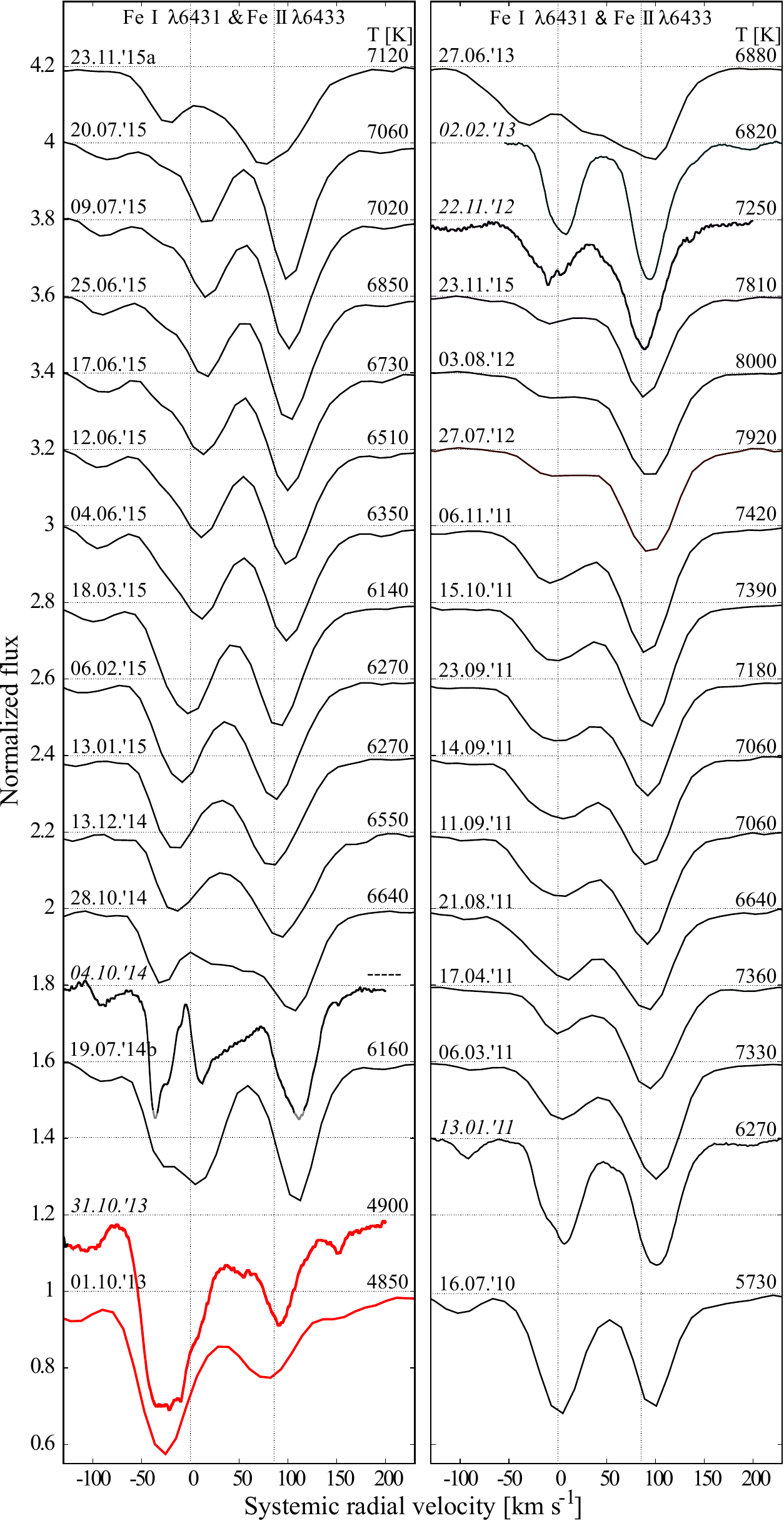}
\caption{Line profile variation of Fe\,{\sc i}
$\lambda$6431 and Fe\,{\sc ii} $\lambda$6433 with their central positions indicated by the vertical
lines. Spectra in red mark the outburst period. Dates listed in italic refer to high-resolution echelle 
spectra. The temperature value obtained from each observation is indicated.}
\label{fig:FeIFeII}
\end{center}
\end{figure}

\begin{table}
\centering
\caption{Effective temperatures. Errors are $\pm$130\,K.}
\label{tab:EW-Teff}
\begin{tabular}{cccc}
\hline 
Date & JD & $T_{\rm eff}$ & Note\\
& (d) & (K) & \\
\hline 
2010-07-16 & 5394.5 &  5730   &   \\
2010-09-24 & 5464.4 &  ---  & (1) \\
2011-01-13 & 5574.6 &  6270   &   \\
2011-03-06 & 5627.2 &  7330   &   \\
2011-04-17 & 5669.6 &  7360   &   \\
2011-08-21 & 5795.4 &  6640   &   \\
2011-09-04 & 5809.4 &  6910   &   \\
2011-09-11 & 5816.4 &  7060   &   \\
2011-09-14 & 5819.6 &  7060   &   \\
2011-09-14 & 5819.4 &  ---  & (1) \\
2011-09-23 & 5828.5 &  7180   &   \\
2011-10-15 & 5850.5 &  7390   &   \\
2011-11-06 & 5872.2 &  7420   &   \\
2012-07-27 & 6136.6 &  7920   &   \\
2012-08-03 & 6143.4 &  8000   &   \\
2012-09-28 & 6199.4 &  7810   &   \\
2012-11-22 & 6254.4 &  7250   &  \\
2013-02-02 & 6326.3 &  6820   &  \\
2013-06-27 & 6471.5 &  6880   &  \\
2013-10-01 & 6567.4 &  4850   &  \\
2013-10-31 & 6597.4 &  4900   &  \\
2014-07-18 & 6857.6 &  6160   &  \\
2014-07-19 & 6858.5 &  6160   &  \\
2014-10-01 & 6931.6 &  ---  & (1) \\
2014-10-04 & 6935.4 &  ---  & (2) \\
2014-10-28 & 6959.3 &  6640   &  \\
2014-12-13 & 7005.3 &  6550   &  \\
2015-01-13 & 7036.2 &  6270   &   \\
2015-02-06 & 7060.2 &  6270   &   \\
2015-03-18 & 7100.6 &  6140   &   \\
2015-06-04 & 7178.5 &  6350   &   \\
2015-06-12 & 7186.5 &  6510   &  \\
2015-06-17 & 7191.4 &  6730   &  \\
2015-06-25 & 7199.6 &  6850   &  \\
2015-07-09 & 7213.5 &  7020   &   \\
2015-07-20 & 7224.4 &  7060   &  \\
2015-11-23 & 7350.5 &  7120   &  \\
2015-11-24 & 7351.4 &  7030   & \\ 
\hline
\end{tabular}

\smallskip
(1) SAO spectrum with gap at 6430\AA. \\
(2) Strong emission in Fe\,{\sc i} $\lambda$6430.

\end{table}

The drop in brightness and the change in spectral appearance during the outburst are clear signs for a
decrease in effective temperature of the star.

In the spectra of cool supergiants the lines of Fe\,{\sc i} and Fe\,{\sc ii} have been found to be 
valuable temperature tracers, because their line strengths change in opposite directions for small 
variations in effective temperature. We utilize this property and estimate the temporal temperature 
variation of $\rho$\,Cas over the full observing period using the 
Fe\,{\sc i}\,$\lambda$6431/Fe\,{\sc ii}\,$\lambda$6433 line depth ratio, calibrated by high resolution 
spectra (Elodie and UVES POP) of 19 late-A to early-K supergiants with precise $T_{\rm eff}$ values from 
the literature \citep{2007MNRAS.378..617K}. These data were supplemented with two hotter A-type stars 
with effective temperatures according to their spectral class \citep{2010ApJS..186..191C}. This 
calibration provides temperature estimates with an error of $\pm$130\,K. The profiles of the two 
adjacent lines are shown in Figure \ref{fig:FeIFeII}, and the temperature values obtained from our 
medium- and high-resolution observations are given in Table \ref{tab:EW-Teff}. For certain epochs no 
temperature could be derived due to gaps in the echelle spectra of NES or during the phase when the 
profiles display strong pollution with a pronounced emission component, as observed for example on 2014 
October 4. The temperature variation is shown in the middle panel of Figure \ref{fig:photom}. 

As \citet{2014ARep...58..101K} have also observed $\rho$\,Cas back in 2010--2011 and determined effective 
temperatures from their high-resolution spectra, we superimpose their values on our temperature curve 
(blue dots in the middle panel of Figure \ref{fig:photom}) and find that our determinations from  
lower spectral resolution data agree very well with theirs so that we are confident about our 
temperature measurements.

In general, the observed variations in effective temperature follow the trend of the light curve, as was 
also noted by \citet{2003ApJ...583..923L}: the star appears hotter during bright phases and cooler
during fainter phases and during the outburst. However, we note that there are epochs in which the 
correlation between temperature and brightness is less accurate. These are the epochs during the onset 
of the outburst (June 2013) and in the recovery phase (July -- October 2014). Here, the absorption lines 
are strongly distorted due to the increased kinematics within the rapidly expanding atmosphere and due 
to the appearance of superimposed emission. Therefore, we would like to caution that the 
method of the line depth ratio has its limitations. It can provide only reliable temperature values as 
long as none of the absorption lines is saturated or filled/polluted with a significant amount of 
emission. Moreover, we would like to stress that in an object like $\rho$\,Cas with a very extended, 
hence diluted, and highly dynamical atmosphere, the variation in brightness is not solely caused by the 
variation in effective temperature. The global dynamics within the atmosphere of a pulsating star can 
severely influence the optical depth, i.e., the apparent radius of the star, and 
consequently impact the brightness as well. It is therefore not surprising that our temperature 
estimates during the outburst imply an equally cool temperature for $\rho$\,Cas as during the millenium 
outburst, which is verified by the appearance of TiO band absorption, although the $V$ brightness 
dropped by only half the value.

\subsection{Atmospheric dynamics}

The outburst is expected to be accompanied by an enhanced atmospheric dynamics. 
In stars with very extended atmospheres such as $\rho$\,Cas the formation of individual 
spectral lines is determined by the contribution function reaching maxima in different regions of the 
atmosphere. The locations of these maxima depend on the excitation energies of the individual lines. 
Hence, lines with diverse excitation energies can yield different radial velocity values, because the 
contributions to their emergent profiles can originate from various portions along the atmospheric 
height. These formation regions are typically associated with different velocities within a pulsating 
atmosphere.

To study the vertical velocity structure within the extended, pulsating photosphere of $\rho$\,Cas, we 
selected several strategic absorption lines of different elements in various ionization stages and 
covering a large range in excitation energies. We measured the midpoint positions of the full line width 
at half maximum and computed from it the center-of-gravity radial velocity of each absorption line 
associated with the mean line formation region. This method minimizes the influence of the wind, which 
manifests itself in the profiles by extended blue wings. Our measurements are shown for four 
representative lines in the bottom panel of Figure \ref{fig:photom}. Typical sources for errors result 
from the stability of the spectrograph and from the adjustment of the continuum. We estimate that the 
total error in velocity will not exceed $\pm 1$\,km\,s$^{-1}$. This errorbar is included in the bottom 
left of the plot.

In general, we observe that all lines display radial velocity variations. The low-excitation lines of 
La\,{\sc ii} $\lambda$6390 and Fe\,{\sc i} $\lambda$6412 with excitation energies of 0.3\,eV and 
2.5\,eV, respectively, are most sensitive to conditions in the uppermost layers of the photosphere.  
During pre-outburst (i.e. quiescence) phase, these lines have usually the highest amplitudes with total 
values of 15--17\,km\,s$^{-1}$. The two high-excitation lines, Si\,{\sc ii} $\lambda$6347 (8.12\,eV) and 
N\,{\sc i} $\lambda$8629 (10.69\,eV), trace the kinematics within the deeper photosphere, and their 
amplitude is with about 9--10\,km\,s$^{-1}$ slightly lower. We note, however, that our observing cadence 
is rather coarse so that these amplitudes can only be considered as rough estimates and consequently 
present lower limits to the real kinematics during quiescence. In fact, much tighter sampling resulted 
in slightly higher amplitude values \citep[e.g.,][]{1986PASP...98..914S, 1998A&A...330..659L, 
2003ApJ...583..923L, 2014ARep...58..101K} but maintaining the general trend of higher amplitudes with
lower excitation energy.

Immediately before, during and after the outburst the amplitudes of all lines are greatly enhanced. 
Particularly interesting is the strong blueshift and the formation of an additional high-velocity blue 
wing in the low-excitation lines in June 2013 (see Fig.\,\ref{fig:app1}) reaching out to about 
$-$110\,km\,s$^{-1}$, hence indicating that rapid expansion of the outermost layer has started already 
when the brightness was still close to maximum. Such a phase lag between brightness and radial velocity 
was also found by \citet{2003ApJ...583..923L}.

The high radial velocity amplitudes observed more than one year after the outburst \citep[i.e., after 
the recovery of the brightness, see also the recent work by][]{2018ARep...62..623K} might indicate that 
the atmosphere itself did not yet fully settle back to its equilibrium, i.e., quiescence state but is 
still strongly oscillating. Support for such an interpretation is provided by the high-excitation lines 
forming generally much deeper in the atmosphere. These lines also show large velocity displacements (see 
Figs.\,\ref{fig:app2} and \ref{fig:app3}), in agreement with a highly dynamical and disordered 
photosphere.

\subsection{Emission lines}

Another noteworthy detected property is the emergence of emission lines, and most interesting is hereby 
the emission in the line Fe{\sc i}\,$\lambda$6359. It is observed in our post-outburst spectra in 2014 
as well as during phases of maximum brightness epochs within the quiescence phase of $\rho$\,Cas 
(Fig.\,\ref{fig:app5}). Fe{\sc i}\,$\lambda$6359  belongs to the `splitting sensitive' group of lines 
\citep[see][for the discovery and definition of line splitting in the spectrum of 
$\rho$\,Cas]{1961ApJ...134..142S, 1998A&A...330..659L, 2003ApJ...583..923L, 2014ARep...58..101K} 
based on its low excitation potential (0.86\,eV).  

In contrast to this occasional appearance of emission, \citet{1998A&A...330..659L} and 
\citet{2006ApJ...651.1130G} noted that an emission component in the forbidden line [Ca{\sc ii}] 
$\lambda$7324 is always present. As this line appeared to be static and hence circumstellar in origin, 
its position in the spectra at $-$47\,km\,s$^{-1}$ has been used to determine the systemic radial 
velocity of $\rho$\,Cas.

The forbidden lines of [Ca{\sc ii}] $\lambda\lambda$7291,7324 are also seen in our red spectra
(middle panel of Fig.\,\ref{fig:spec}). Inspection of our time series reveals that the shape and 
strength of their profiles strongly depends on the interplay between the static emission and the 
underlying dynamic (in position, shape and strength) absorption profile (Fig.\,\ref{fig:app5}). In 
particular in the first half of 2015 the emission is basically completely compensated by deep and 
broad underlying absorption components.

\begin{figure}
\begin{center}
\includegraphics[width=0.9\hsize,angle=0]{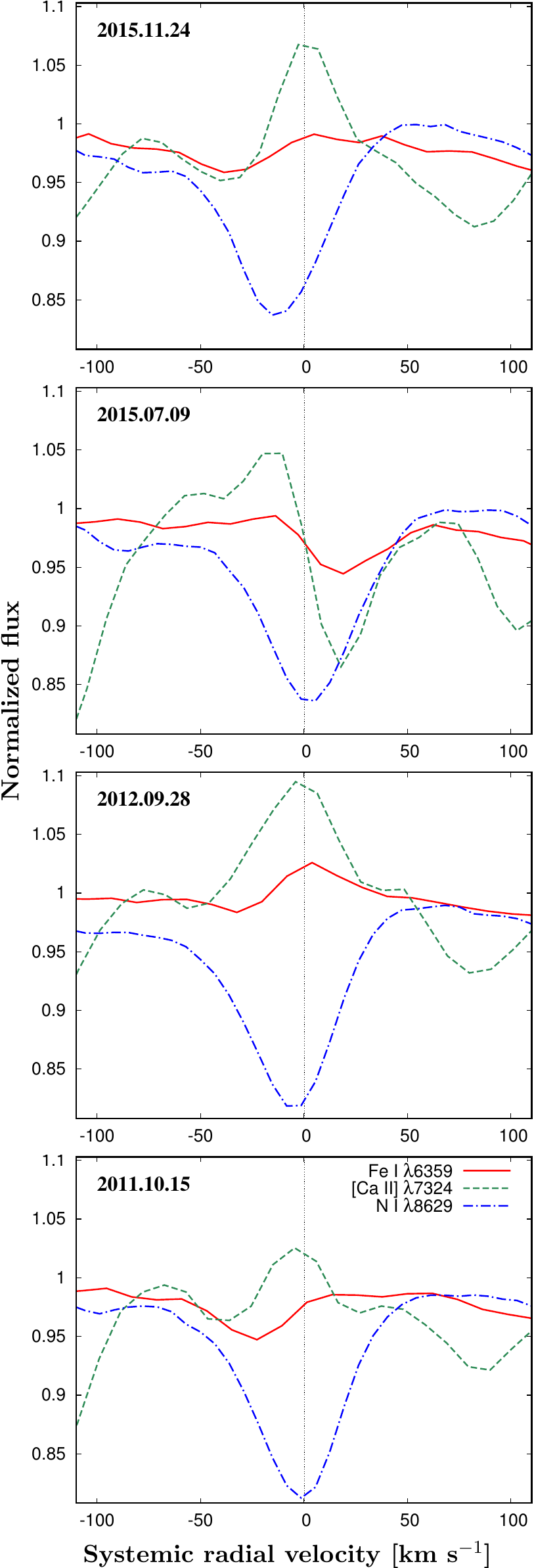}
\caption{Comparison of line profiles in four selected epochs. The emission component in 
Fe{\sc i}\,$\lambda$6359 varies synchronously with [Ca{\sc ii}] $\lambda$7324 implying static emission 
in the low-excitation iron line, which is not seen in the high-excitation N{\sc i} $\lambda$8629 line.}
\label{fig:epochs}
\end{center}
\end{figure}

Examination of the position of the emission component in the Fe{\sc i}\,$\lambda$6359 line in 
comparison to [Ca{\sc ii}] $\lambda$7324 shows that a tight correlation exists in the behaviour of both 
lines. This is demonstrated over the full time series shown in Fig.\,\ref{fig:app5} and highlighted 
in Fig.\,\ref{fig:epochs} where we display the profiles of [Ca{\sc ii}] $\lambda$7324 and
Fe{\sc i}\,$\lambda$6359 in four selected epochs (from top to bottom): the 
most blue-shifted, the most red-shifted, and two intermediate ones. Obviously, the emission bumps in 
Fe{\sc i}\,$\lambda$6359 are in phase with those of [Ca{\sc ii}] $\lambda$7324.
Therefore, we conclude that the profile of Fe{\sc i}\,$\lambda$6359 
also contains a static circumstellar emission component, exactly as it is the case for [Ca{\sc ii}]. 
This static emission component disturbs the absorption profile and results in a permanent `splitting' of 
the line, although it is not always prominent or obvious in the profiles.

\citet{1961ApJ...134..142S} has found that splitting occurs in lines with an excitation potential up to 
2.9\,eV of their lower levels. Many of these potentially split lines actually behave similarly to 
Fe{\sc i}\,$\lambda$6359. Depending on their strength and sensitivity to temperature changes, they  
exhibit the splitting only at certain epochs. For instance, several absorption lines of Fe{\sc i} in our 
spectral time series are split at just one epoch (2014-10-04) while during the remaining observation 
period they display pure absorption profiles albeit varying with moderate asymmetry. Also for these 
lines, we can recognize that the asymmetry within the line profiles can be ascribed to a static 
circumstellar emission component in line (and in phase) with the emission of [Ca{\sc ii}] $\lambda$7324.
Likewise we might interpret the asymmetries seen in Fe{\sc i} lines with excitation potentials higher 
than 2.9\,eV with circumstellar emission (see Fe{\sc i} $\lambda$6400 with EP = 3.60\,eV in 
Fig.\,\ref{fig:app1}) because they all follow the same trend as the Fe{\sc i}\,$\lambda$6359 line.

When inspecting lines with much higher excitation potential (see, e.g., Si{\sc ii} $\lambda$6347 
with EP = 8.12\,eV and N{\sc i} $\lambda$8629 with EP = 10.69\,eV in Fig.\,\ref{fig:app2} and 
Fig.\,\ref{fig:app3}, respectively) their absorption profiles display no obvious indication for 
superimposed circumstellar emission. For demonstration purposes we included the profile of N{\sc i} 
$\lambda$8629 in Fig.\,\ref{fig:epochs}. In contrast to the low-excitation lines that are contaminated 
with circumstellar emission, this high-excitation line displays radial velocity shifts around the 
systemic velocity together with a mild wiggling of its profile shape, in agreement with pulsation
movements in the atmosphere of $\rho$\,Cas.

Finally, it should be mentioned that H$\alpha$ exhibits time variable emission components in both line 
wings (Fig.\,\ref{fig:app4}). This was also noted by \citet{2006ApJ...651.1130G}. 
However, we want to stress that the emission variability seen in H$\alpha$ is not in phase
with [Ca{\sc ii}] $\lambda$7324, rendering it less likely that this emission is of static circumstellar 
origin. Instead, H$\alpha$ forms over a large volume of the stellar photosphere extending into 
the base of the wind where the conditions in terms of density and temperature are still favourable 
to partially ionize hydrogen and to generate recombination line emission in a measurable amount.

\section{Discussion}\label{sec:disc}

\subsection{Circumstellar material}

\begin{figure*}
\begin{center}
\includegraphics[width=0.49\hsize,angle=0]{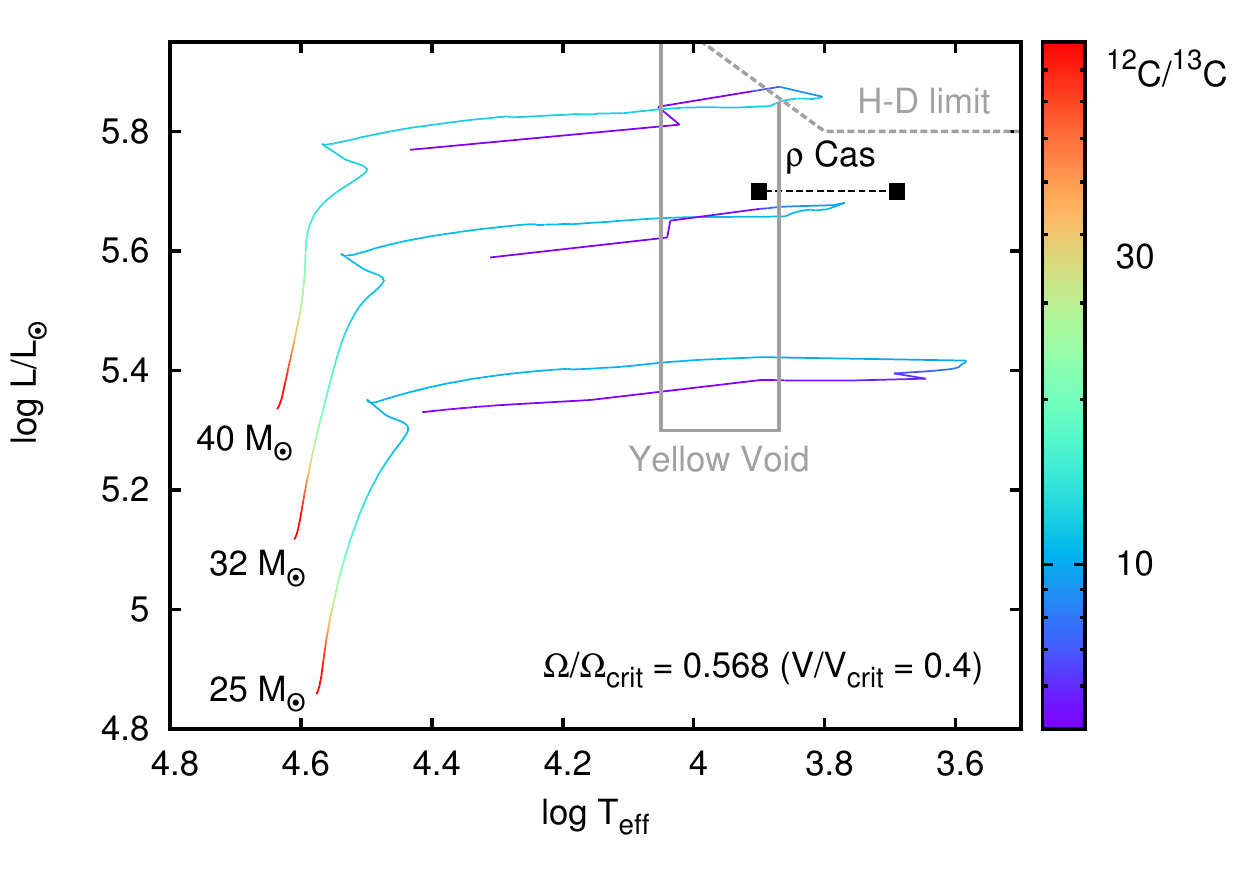}
\includegraphics[width=0.49\hsize,angle=0]{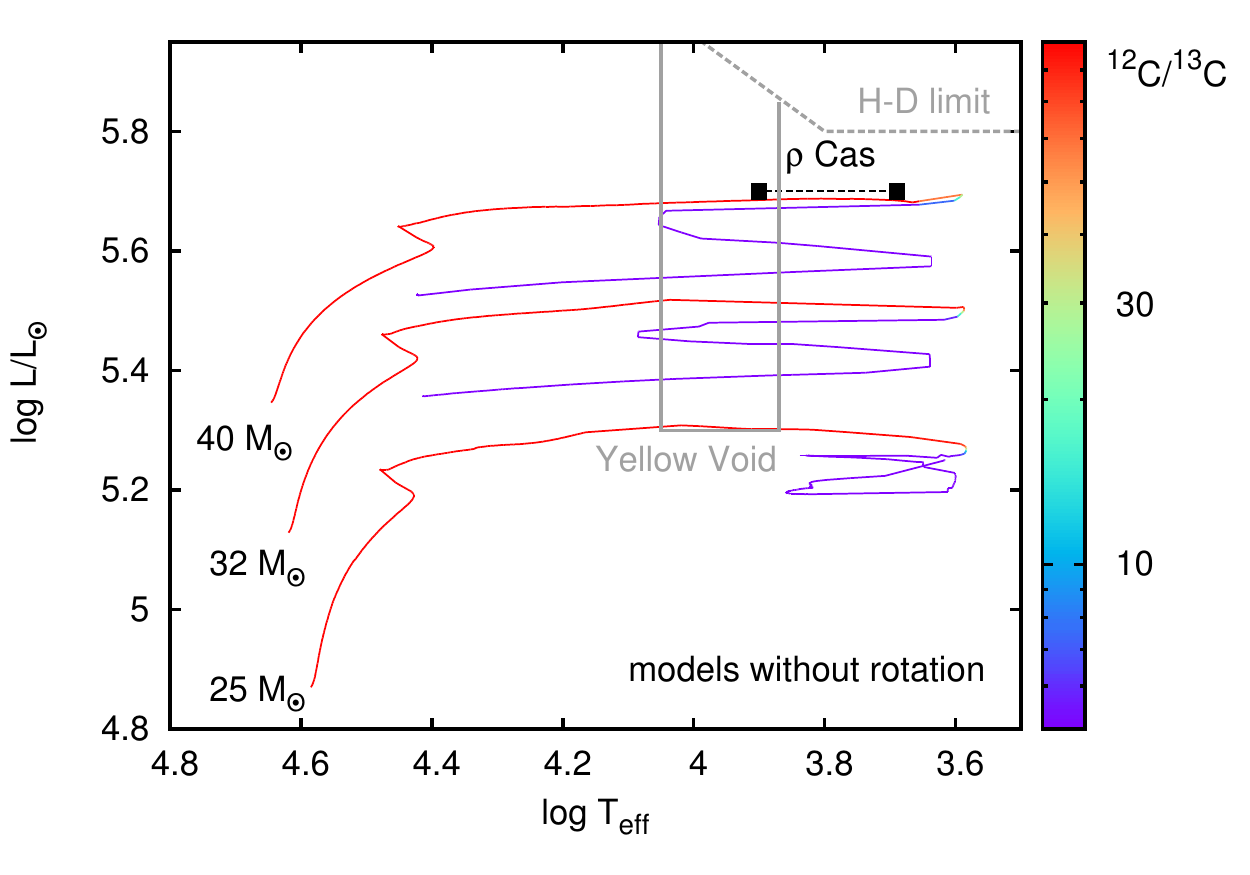}
\caption{Position of $\rho$\,Cas in the HR diagram along with Geneva stellar evolutionary tracks for
solar metallicity models with (left panel) and without (right panel) rotation 
\citep[from][]{2012A&A...537A.146E}. The luminosity of $\rho$\,Cas is from \citet{1978ApJS...38..309H}. The 
temperature spreads from the hot, quiescent to the cool, outburst state (Table\,\ref{tab:EW-Teff}). 
The positions of the Yellow Void instability region \citep{1997MNRAS.290L..50D} and 
the Humphreys-Davidson limit \citep{1994PASP..106.1025H} are shown in gray. The color coding (in logarithmic 
scale) refers to the surface abundance ratio of $^{12}$C/$^{13}$C.}
\label{fig:HRD-C}
\end{center}
\end{figure*}

In general, emission lines of [Ca{\sc ii}] are a reasonable tracer for dense, circumstellar gas around 
hot \citep[e.g.,][]{2010A&A...517A..30K, 2012MNRAS.423..284A, 2018MNRAS.tmp.1673M} and cool stars 
\citep{2013ApJ...773...46H, 2017ApJ...836...64H, 2017ASPC..508..239A, 2017ASPC..510..162A}. Our findings 
of static circumstellar emission in basically all low to medium-excitation lines in the spectra of 
$\rho$\,Cas with the same spectroscopic appearance as the [Ca{\sc ii}] lines hence reinforces, that the 
star must be surrounded by a significant amount of gas, possibly confined in a stable shell or ring.

Further support for warm and dense circumstellar material is provided by molecular emission, primarily
by CO band emission. These bands have been detected in the near-infrared spectra ($> 2.3\,\mu$m) of many 
B[e] supergiants, in which the emission typically originates from a circumstellar ring or the inner edge 
of a molecular disc \citep{2000A&A...362..158K, 2016A&A...593A.112K, 2017AJ....154..186K, 
2010MNRAS.408L...6L, 2014MNRAS.443..947L, 2012A&A...548A..72C, 2012A&A...543A..77W, 2012MNRAS.426L..56O, 
2013A&A...558A..17O, 2015AJ....149...13M, 2018A&A...612A.113T, 2018MNRAS.tmp.1673M}. In $\rho$\,Cas, CO 
band emission was first reported by \citet*{1981ApJ...248..638L}. The excitation of these bands requires 
a temperature range of 2000--5000\,K which is much lower than the stellar effective temperature. 
Monitoring of the CO bands revealed that they switch from pure emission to pure absorption and back 
\citep{2006ApJ...651.1130G, 2007PASJ...59..973Y}. Interestingly, the phases of pure and intense emission 
were found to coincide with phases of atomic line emission, which agree with phases of maximum light. 

According to \citet{2006ApJ...651.1130G} the CO band emission is centred on the systemic velocity and 
displays only mild radial velocity variations, whereas the absorption can appear significantly red or 
blue-shifted (up to 30\,km\,s$^{-1}$). Based on these findings we propose that the CO band emission 
itself is always present in the spectra, just as the [Ca{\sc ii}] emission, and likewise originates 
from the circumstellar gas, while the absorption forms in the outermost atmospheric layer. 

Considering that during the atmospheric expansion caused by the pulsation activity of the hypergiant 
star the temperature within its very outer layer can be considerably lower than the effective 
temperature, and hence also lower than the dissociation temperature of CO which is around 5000\,K, the 
conditions for the formation of CO molecules can be met. These hot molecules will absorb the continuum 
emission from the star in the near-infrared, causing CO band absorption. With further expansion hence 
cooling, the absorption bands superimposing the circumstellar emission will first compensate the 
emission and then continue to grow and dominate the near-infrared spectral appearance until the 
expansion stops and contraction sets in, reversing the process. During phases of maximum light the star 
is hottest and therefore most compact. Even its outermost layers are too hot for CO molecules to exist. 
Therefore, no absorption takes place and only the circumstellar emission is observable. 

The line profiles of the circumstellar emission components are very narrow and hence do not provide
information about the dynamics within their formation region. For instance, the profiles of the 
[Ca{\sc ii}] lines, which in our spectra are narrow and single-peaked, display no indication 
for kinematical broadening beyond our spectral resolution. The same holds for the stable circumstellar 
emission component of the Fe{\sc i} $\lambda$5328 line measured by \citet{1998A&ARv...8..145D} in their 
high-resolution spectra. If $\rho$\,Cas is surrounded by a Keplerian rotating ring (or disc), as was 
proposed to be the case for the two hotter YHGs IRC\,+10420 \citep{2007ApJ...671.2059D} and V509\,Cas 
\citep{2017ASPC..510..162A}, the orientation of the $\rho$\,Cas system would have to be (close to) 
pole-on. 

So far, the origin of the static circumstellar gas is unclear. It might be the remnant of one of the 
previous mass ejection, i.e. outburst events, or even of a possible previous RSG mass-loss.
\citet{1998A&ARv...8..145D} proposed that the material confinement could be caused by a stationary 
shock at the interface of the interstellar medium and the stellar wind. In this scenario, the stellar 
wind streams with high velocity into the shock and leaves it with very low velocity, explaining the 
stable position of the emission and the narrowness of the lines \citep[see Chap. 4, p. 43 
in][]{1997PhDT........29L}. Alternatively, depending on the pressure of the interstellar medium, the 
interface between the stellar wind and the interstellar medium might take 
the form of a stagnation region determined by a stagnation point, which is known as the classical scenario of an 
astrosphere \citep[e.g.,][]{2006A&A...454..797N, 2014ASTRP...1...51N}. Along the stagnation line 
(passing through the stagnation point) and all other streamlines in its vicinity, the speed of 
the matter reduces to (almost) zero velocity on both sides in the stagnation region, implying an 
extremely long arrival time for the material from the star.

In any case, abundance studies of the circumstellar matter in relation with a closer look at the 
evolutionary state of $\rho$\,Cas might help to discriminate between the various scenarios.

\begin{figure*}
\begin{center}
\includegraphics[width=0.49\hsize,angle=0]{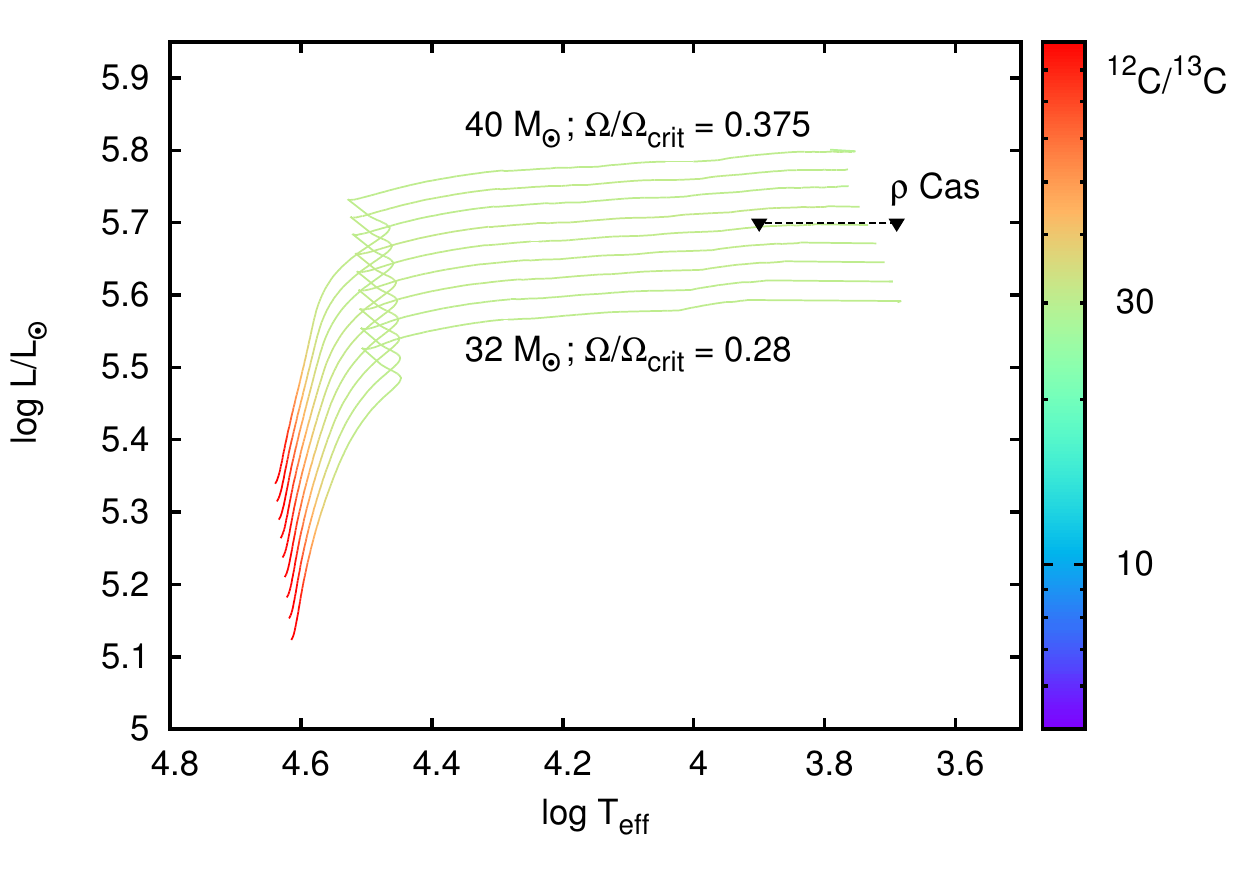}
\includegraphics[width=0.49\hsize,angle=0]{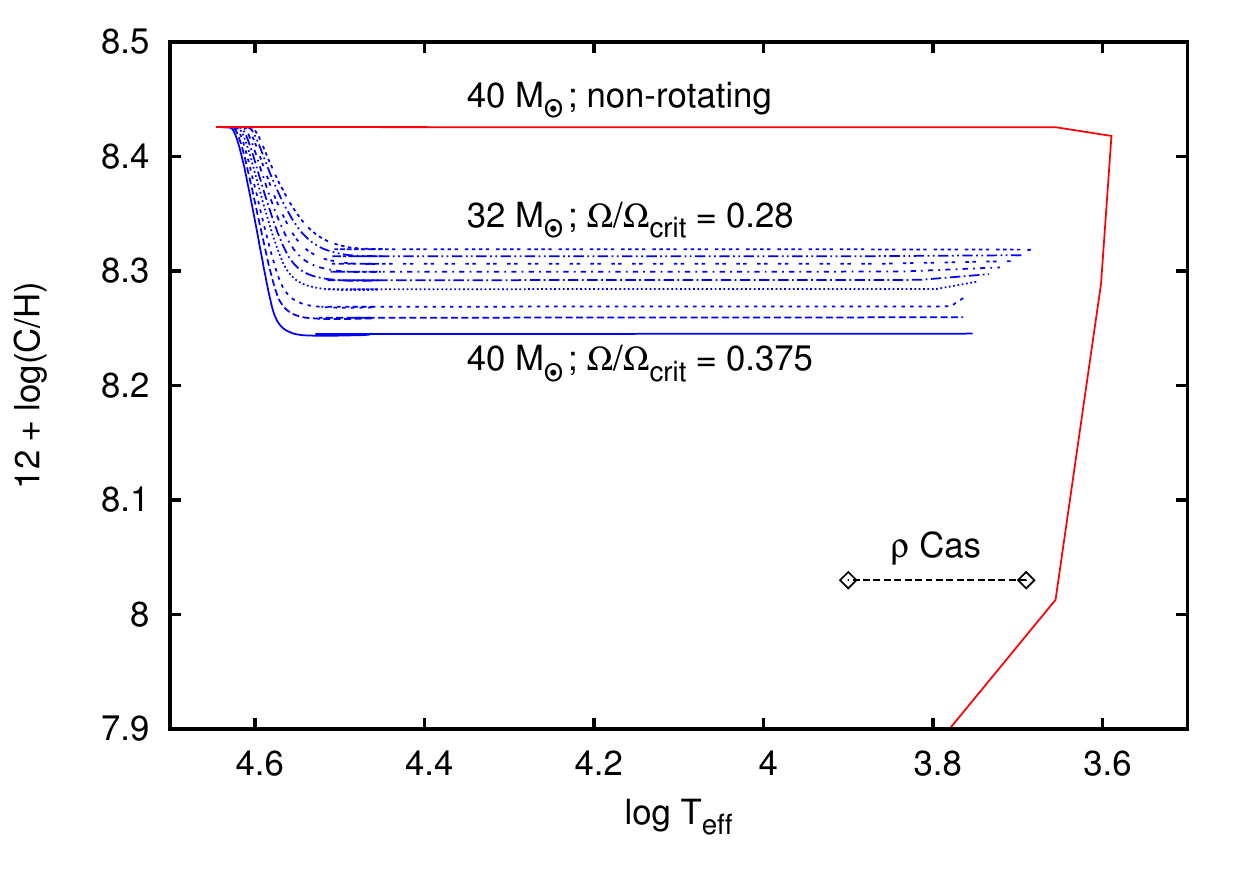}
\caption{Left: Geneva stellar evolutionary tracks up to the RSG state for models between 32 and 
40\,M$_{\odot}$ with rotation. The rotation rate is determined such that the star maintains a 
$^{12}$C/$^{13}$C ratio of 30 from the end of the main-sequence up to the position of $\rho$\,Cas.
Right: Evolution of the surface $^{12}$C abundance for the same parts of the stellar evolution models 
as in the left panel. Included are the results for the 40\,M$_{\odot}$ model without rotation up to the 
blue loop and the measured value of 12$+$log(C/H) = 8.03 of $\rho$\,Cas from 
\citet{1983BCrAO..66..119B}. A conservative error estimate for this value is $\pm$0.1.}
\label{fig:C-abund}
\end{center}
\end{figure*}

\subsection{Evolutionary state of $\rho$\,Cas}

\citet{1988Ap.....28..202B} reported a large sodium overabundance from a non-LTE anaysis of 
optical Na I lines. They measured [Na/H]=0.72 and [Na/Fe]=0.67 (their Table 3). The latter value is the 
largest value for the F-type supergiants they analysed, signaling an excess abundance of sodium in
this Yellow Hypergiant. Alike, the intense emission in 
Na{\sc i} detected in the near-infrared spectra \citep{2006ApJ...651.1130G, 2007PASJ...59..973Y} has 
been assigned to a high Na{\sc i} abundance \citep{1994PASJ...46..395T, 1995ApJ...451..298E}.
Moreover, \citet{1981BCrAO..63...68B} and \citet{1983BCrAO..66..119B} derived a carbon 
abundance of 12$+$log(C/H) = 8.06 and 8.03, respectively, based on measurements of the 
line equivalent widths of six representative carbon lines in the optical spectra of $\rho$\,Cas. 
Based on these sodium abundance measurements, \citet{1998A&ARv...8..145D}
proposed that $\rho$\,Cas must be a post-red supergiant, evolving along the blueward loop in the 
HR-diagram.

Additional information about the possible evolutionary state of $\rho$\,Cas is provided by the carbon 
isotope ratio $^{12}$C/$^{13}$C. While this ratio cannot be directly measured on the stellar surface, 
\citet{2009A&A...494..253K} argued that it translates into the molecular isotope abundance ratio 
$^{12}$CO/$^{13}$CO which can be easily derived from even moderate-resolution near-infrared spectra.
The $^{12}$CO/$^{13}$CO ratio measured within the circumstellar material hence traces the surface 
$^{12}$C/$^{13}$C ratio at the time of mass ejection. Based on the strength of the $^{13}$CO emission 
and absorption features seen in their high-resolution near-infrared spectra \citet{1981ApJ...248..638L} 
derived a carbon isotope ratio of $^{12}$C/$^{13}$C = 30. 

Using the Geneva stellar evolutionary tracks from \citet{2012A&A...537A.146E} for  
significantly rotating ($\Omega/\Omega_{\rm crit}$ = 0.568, corresponding to $V/V_{\rm crit}$ 
= 0.4) and non-rotating stars with solar metallicity, we compute the carbon isotope ratio along the 
tracks. The results are shown in Fig.\,\ref{fig:HRD-C}. The position of $\rho$\,Cas is included in both 
plots. The stellar luminosity is taken from \citet{1978ApJS...38..309H}, and the minimum (during 
outburst) and maximum temperature values follow from our analysis (Table\,\ref{tab:EW-Teff}). In the 
models with significant stellar rotation a carbon isotope ratio of 30 is reached already during 
the main-sequence evolution, whereas in the temperature range of $\rho$\,Cas the ratio is about 10.5 in 
the pre-RSG stage and drops to 6.65 in the post-RSG stage. Along the tracks with no 
rotation the stars keep the initial (interstellar) value of the carbon isotope ratio of $\sim$ 90 from 
the main sequence up to the RSG stage. Only during the RSG phase, the ratio quickly drops and reaches a 
post-RSG value of about 7 within the current temperature range of $\rho$\,Cas. 

Considering these two extremes, there should also exist stellar models with moderate
rotation in which the observed $^{12}$C/$^{13}$C ratio of 30 could be reached during the pre-RSG phase. 
To search for such models, we utilized the interpolation interface provided by the Geneva
group\footnote{https://www.unige.ch/sciences/astro/evolution/en/database/syclist/}. Using the database 
for solar metallicity, we retrieved tracks for the mass range 32 to 40\,M$_{\odot}$ in steps of
1\,M$_{\odot}$ and with rotation rates ($\Omega/\Omega_{\rm crit}$) such that the carbon isotope ratio
of 30$\pm$1 is achieved in the post-main sequence but pre-RSG state. The required rotation rate  
decreases slightly for decreasing stellar mass. These models are shown in the left 
panel of Fig.\,\ref{fig:C-abund}. We furthermore used these models to compute the evolution of the 
carbon abundance on the stellar surface. The results are presented in 
the right panel of Fig.\,\ref{fig:C-abund} together with $\rho$\,Cas and its measured value 
of 12$+$log(C/H) = 8.03 from \citet{1983BCrAO..66..119B}. Assuming that the 
carbon abundance is representative of the surface abundance at the time of the ejection, and considering 
a conservative value of $\pm$0.1 as error to the carbon abundance value, the observed carbon deficiency 
cannot be achieved during the pre-RSG evolution if at the same time the $^{12}$C/$^{13}$C ratio of 30 
should be kept. 

We also inspected model predictions for the evolution of the sodium surface abundance in these mass and 
rotation velocity bins. While sodium abundances are not provided by the Geneva group, the stellar 
evolution tracks of \citet{2011A&A...530A.115B} contain sodium surface abundances along the evolution 
from the main-sequence up to the RSG state for massive stars rotating with a diversity of rates. We 
inspected their models for initial masses of 30, 35, and 40\,M$_{\odot}$ and rotation 
speeds up to $V/V_{\rm crit} \le 0.4$. We found that these models provide a maximum achievable sodium 
enrichment of [Na/H] $<$ 0.3 and [Na/Fe] $<$ 0.25. These values are far below the observed ones in 
$\rho$\,Cas of 0.72 and 0.67, respectively \citep{1988Ap.....28..202B}.

Considering the evolution of a 40\,M$_{\odot}$ star without (or with only very mild) rotation, the 
observed abundances of carbon and its isotope (right panels of Figs.\,\ref{fig:C-abund} and \ref{fig:HRD-C}, respectively) as well as the overabundance of Na can be reached during or beyond the 
RSG evolution of the star. These abundance considerations suggest that $\rho$\,Cas would be in its 
post-RSG (or blue loop) evolutionary phase. In this scenario, the observed $^{12}$C/$^{13}$C ratio of 30 
would correspond to material that was released during a previous RSG stage.

Based on the velocity-luminosity relation for RSGs derived by \citet{2011A&A...526A.156M}, we obtain a 
wind velocity of $\sim$35\,km\,s$^{-1}$ during the RSG state of $\rho$\,Cas. Considering that the 
enhanced mass loss within the RSG phase took place about 10\,000\,yr ago, the material would have 
travelled over a length of roughly 0.36\,pc, which corresponds to an angular distance of 
$\sim$24\arcsec \ at the distance of 3.1\,kpc for $\rho$\,Cas \citep{2003ApJ...583..923L}. This angular 
distance is too small to be seen on, e.g., WISE images, on which $\rho$\,Cas is saturated. This distance 
is also much smaller than the size of a regular wind-blown bubble resulting from the previous blue 
supergiant phase, so that we can exclude that the material is located within the astrosphere or shock 
region with the interstellar medium. On the other hand, this distance is too large to be 
seen on the HST images presented by \citet{2006AJ....131..603S}, and too large to guarantee that the 
material is dense and hot enough to produce measurable amounts of CO band emission. Therefore, we 
believe that the post-RSG wind, which is faster but much less massive, interacts with the material 
released during the RSG phase. This interaction can lead to a compression and heating of the old 
material (with the appropriate carbon isotope ratio)  and hence give rise to the observed, static CO 
band emission. The temperature of the CO gas is much higher than the dust sublimation 
temperature. In the region of CO band emission, the physical parameters within the environment might 
hence prevent the efficient formation of dust, which could be another reason for the lack of a 
detectable dusty nebula around $\rho$\,Cas.

\begin{figure}
\begin{center}
\includegraphics[width=\hsize,angle=0]{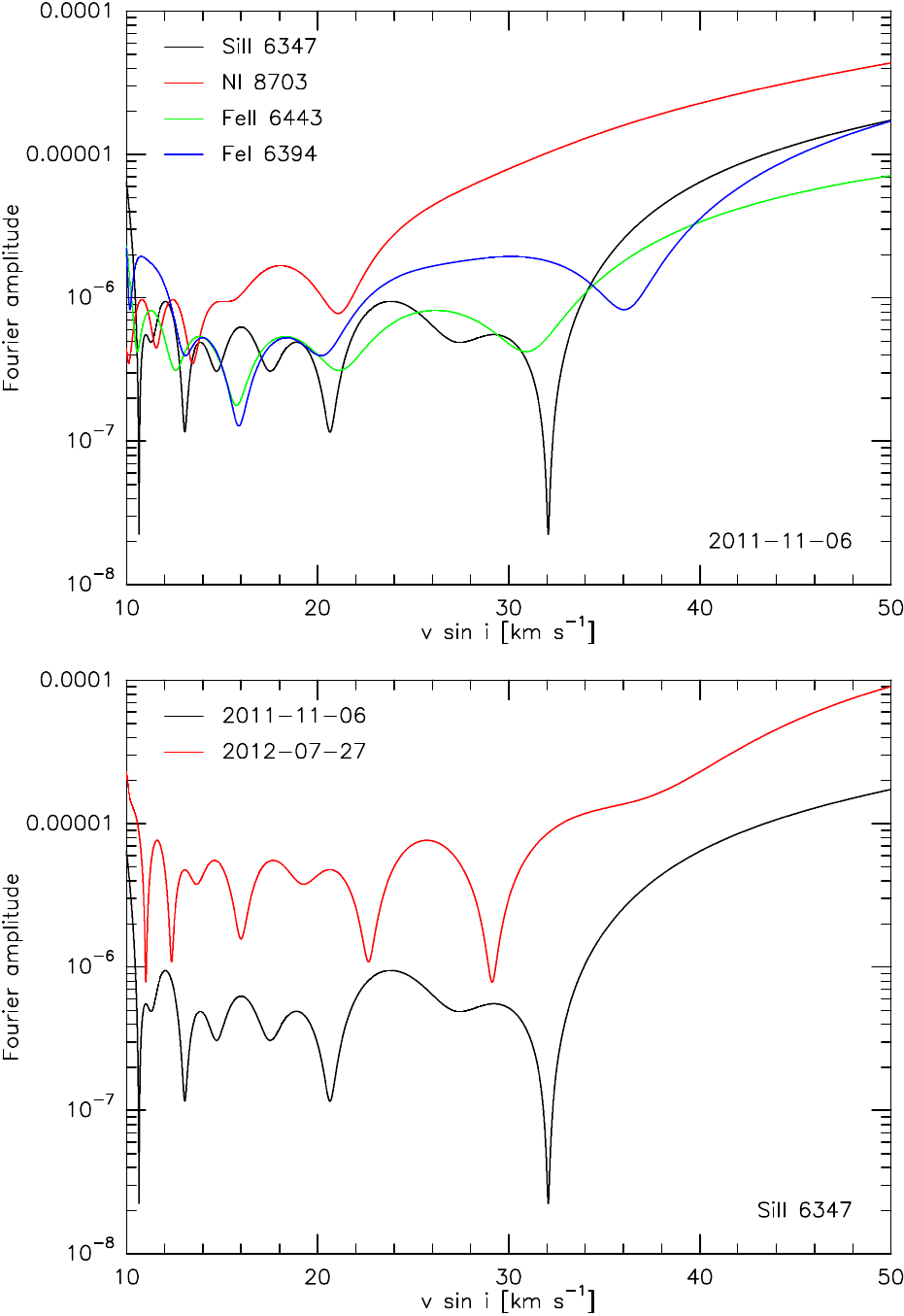} 
\caption{Results from the Fourier transformation of selected lines with apparently
symmetric profile shapes. Top: different lines within the same night; bottom: same line within
two different nights.}
\label{fig:vsini}
\end{center}
\end{figure}

For its spectral type, and considering that $\rho$\,Cas is presumably a very slow rotator, it displays 
extremely broad absorption lines. These were suggested to be broadened by macroturbulence. This type of 
line broadening is typically found in early-type (OB) supergiants in which pulsations play a significant 
role together with stellar rotation \citep[see, e.g.,][]{2002MNRAS.336..577R, 2010ApJ...720L.174S, 
2009A&A...508..409A, 2015A&A...581A..75K}, and is one of the criteria to discriminate YHGs from YSGs 
\citep{1998A&ARv...8..145D}. While $\rho$\,Cas is known to pulsate, not much is known about its real 
rotation velocity, although \citet{1998A&A...330..659L} derived a value of $v\sin i \simeq 25$\,km\,s$^{-1}$ from line-profile fitting.
 
To investigate the possible contribution of rotation to the line profile broadening of $\rho$\,Cas,
we analysed photospheric lines using the method of Fourier transformation. This method only works
properly for symmetric profiles. However, symmetric line profile shapes are rather rare in our data
of $\rho$\,Cas and basically only seen in two nights. Also, given the spectral resolution
of our data, the Fourier transformation can recover only rotation velocities projected to the 
line-of-sight higher than $v \sin i \sim 15$\,km\,s$^{-1}$ with high confidence. The results from our 
analysis are shown in Fig.\,\ref{fig:vsini} for different elements within the same night (top panel) and 
for the same element in the two different nights (bottom panel). If a rotation component projected to 
the line of sight higher than 15\,km\,s$^{-1}$ would be present in these profiles, the zero points of
all elements should indicate the same value of $v \sin i$. However, our data provide no 
consistent value for the rotation velocity. Therefore, we conclude that the line broadening seen in 
$\rho$\,Cas is primarily (maybe even exclusively) due to the large-scale atmospheric dynamics controlled 
by pulsations.

\section{Conclusions} \label{sec:concl}

Based on long-term photometric and spectroscopic monitoring of the YHG $\rho$\,Cas we found that it 
underwent a new outburst in 2013 with a temperature decrease of $\sim$3000\,K and a drop of 
$\sim$0.6\,mag in brightness. The large variability in radial velocities of basically all photospheric 
lines even after the recovery of the stellar brightness indicates that the atmosphere of $\rho$\,Cas
is still far from being back in equilibrium.

Our data indicate that basically all low- to medium-excitation lines display a static emission component 
that displays the same behaviour as the circumstellar [Ca{\sc ii}] emission lines. Consequently, we 
propose that the emission features seen especially during the hot, maximum light phases of $\rho$\,Cas 
are all circumstellar in nature, including the emission in CO bands. Based on an analysis of the 
$^{12}$C/$^{13}$C abundance ratio, we further propose that this circumstellar material is the remnant
from strongly enhanced mass-loss during the previous RSG state. This material is possibly compressed and 
heated by the subsequent post-RSG wind. But its distance from the star is too small to be visible on 
currently available infrared images due to saturation issues. 

Follow-up monitoring of $\rho$\,Cas is indispensable to study in more detail the global dynamics of its
atmosphere, to analyse its pulsation activity, and to identify new outburst phases, which will help
the star to reach again stable atmospheric conditions on its way through the Yellow Void.

\section*{Acknowledgments}

We thank the technical staff at the Ond\v{r}ejov Observatory for the support during the observations. 
This research made use of the NASA Astrophysics Data System (ADS) and of the SIMBAD database, 
operated at CDS, Strasbourg, France. MK, AA, and DHN acknowledge financial support from GA\,\v{C}R 
(grant number 17-02337S). IK and TE acknowledge financial support from the institutional research 
funding IUT40-1 of the Estonian Ministry of Education and Research, and VGK from the Russian Foundation 
for Basic Research (grant 18-02-00029\,a). The Astronomical Institute Ond\v{r}ejov is 
supported by the project RVO:67985815. This research was also supported by the European Union European 
Regional Development Fund, Project ``Benefits for Estonian Society from Space Research and Application'' 
(KOMEET, 2014-2020.4.01.16-0029). We thank the referees for careful reading of our manuscript.

This work is based on observations collected with the Perek 2-m telescope at Ond\v{r}ejov 
Observatory, Czech Republic, the 6-m telescope at the Special Astrophysical Observatory (SAO), 
Russia, and the Mercator 1.2-m telescope on La Palma. We acknowledge with thanks the variable star 
observations from the AAVSO International Database contributed in particular by W. Vollmann.

\bibliographystyle{mnras}
\bibliography{ms} 



\appendix

\section{Selected line profiles}

\begin{figure*}
\begin{center}
\includegraphics[width=0.85\hsize,angle=0]{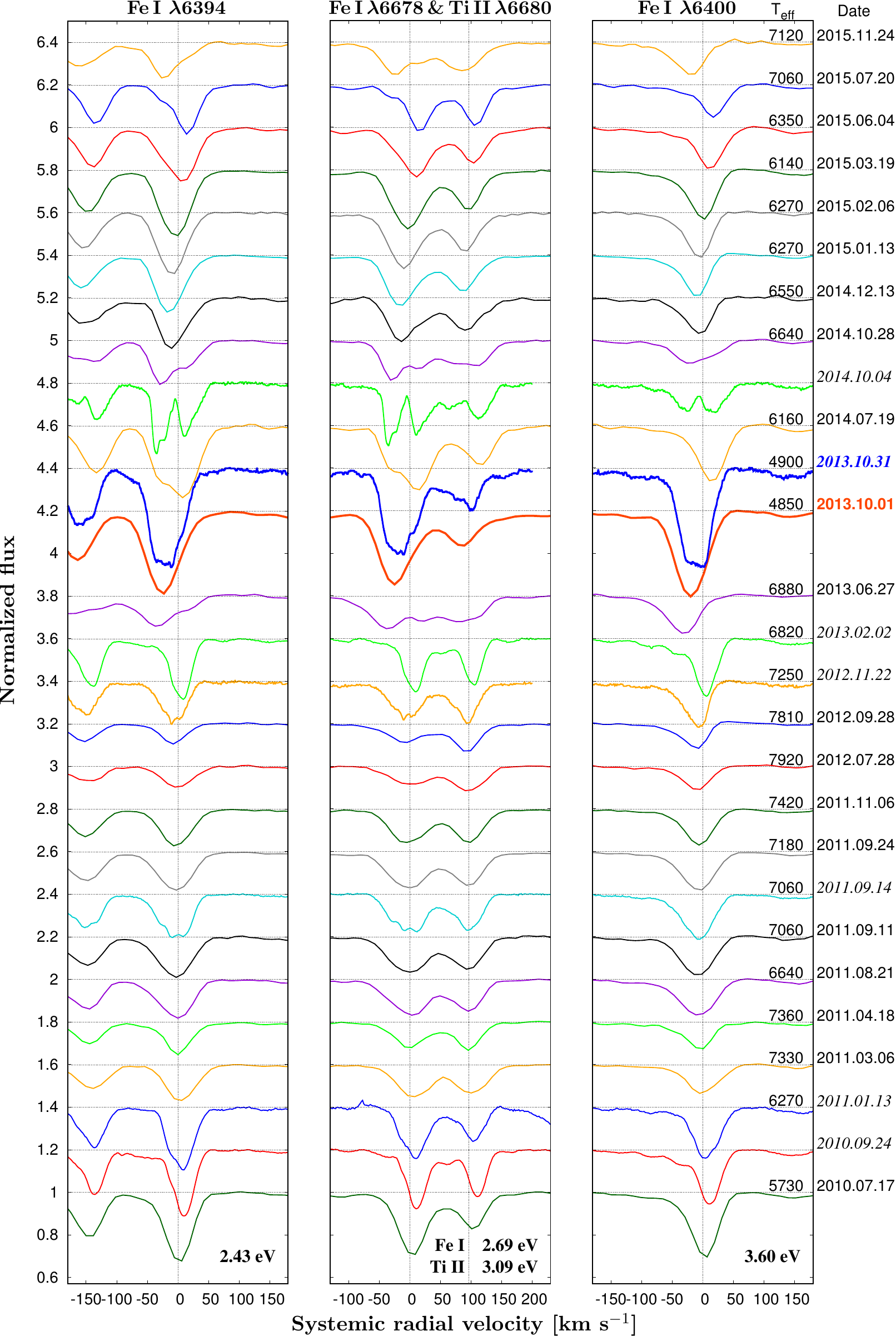}
\caption{Variability in the profiles of selected low-excitation lines in
the years 2010 -- 2015. The excitation potentials are indicated. For easier comparison, the lines from
the same observing date are shown in the same color.}
\label{fig:app1}
\end{center}
\end{figure*}

\begin{figure*}
\begin{center}
\includegraphics[width=0.6\hsize,angle=0]{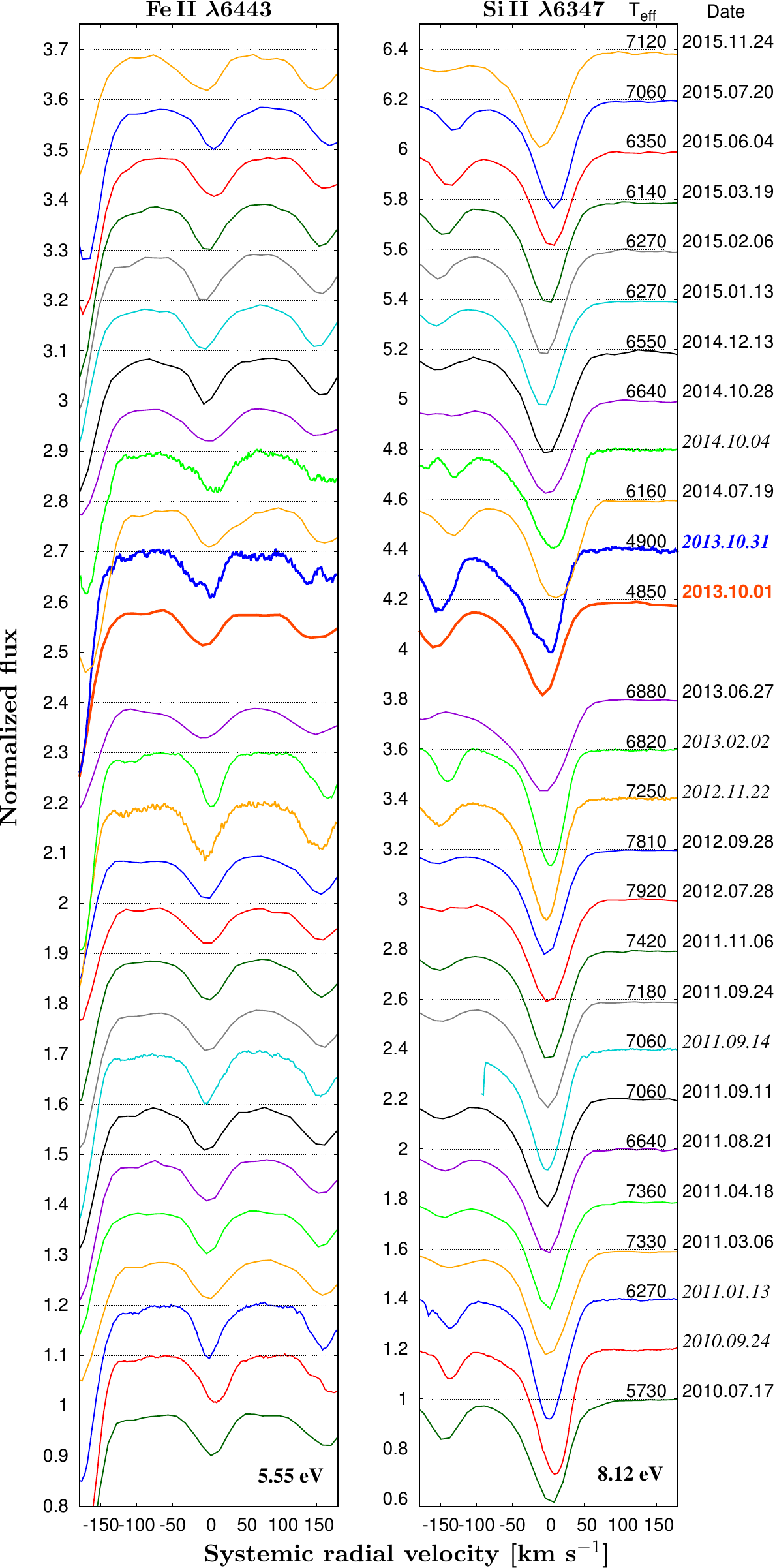}
\caption{As Figure \ref{fig:app1} but for medium-excitation lines.}
\label{fig:app2}
\end{center}
\end{figure*}

\begin{figure*}
\begin{center}
\includegraphics[width=0.6\hsize,angle=0]{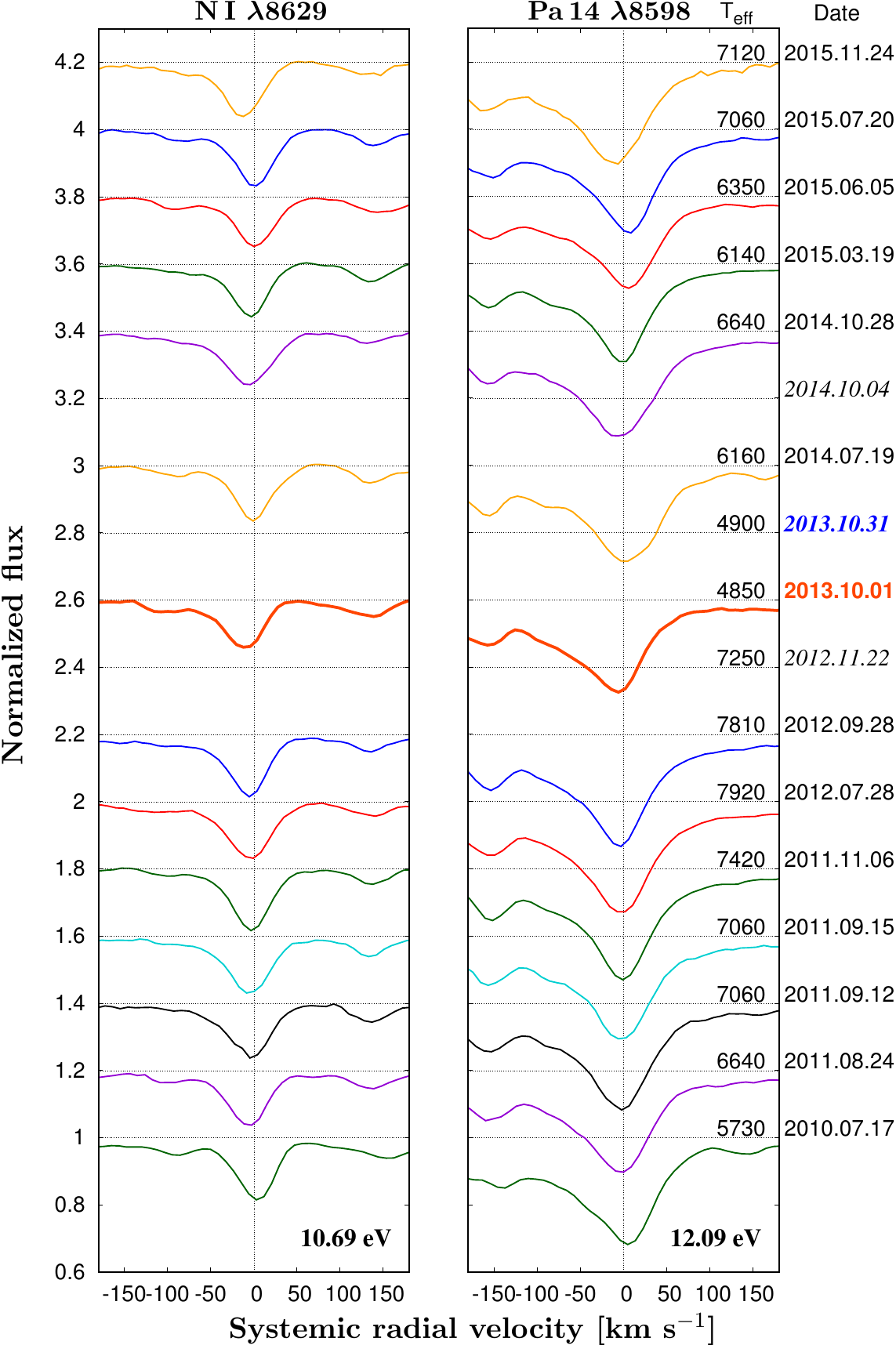}
\caption{As Figure \ref{fig:app1} but for high-excitation lines.}
\label{fig:app3}
\end{center}
\end{figure*}

\begin{figure*}
\begin{center}
\includegraphics[width=0.45\hsize,angle=0]{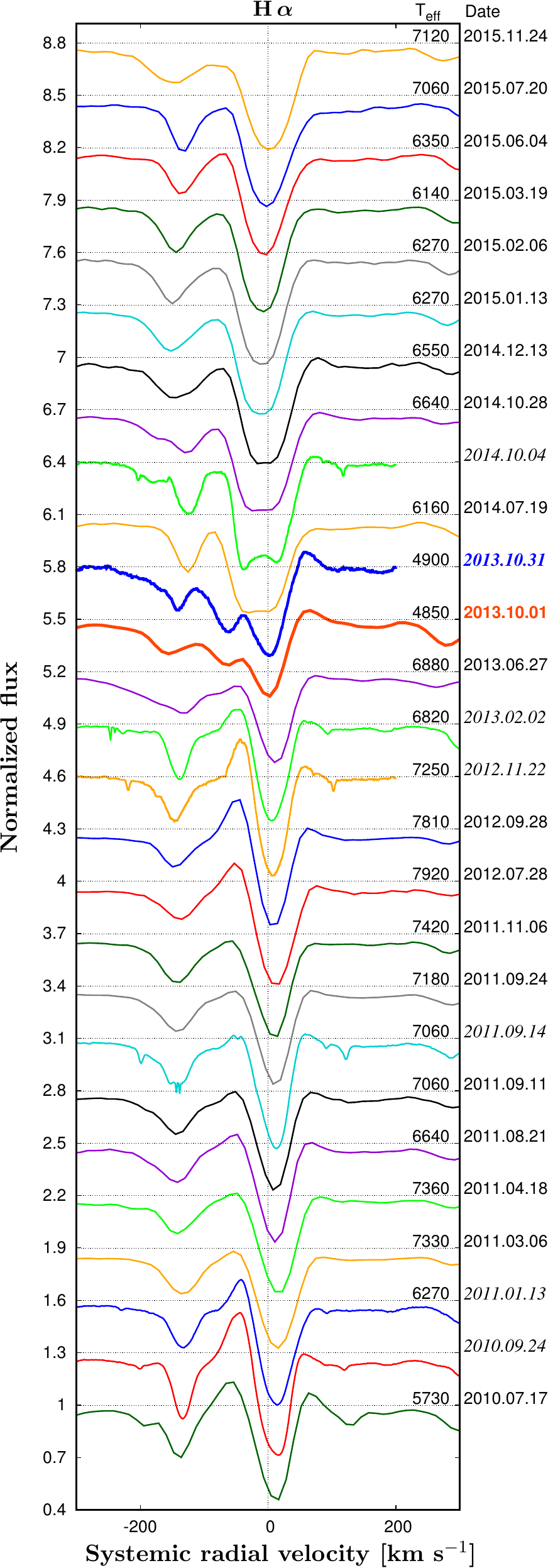}
\caption{Variability in H$\alpha$.}
\label{fig:app4}
\end{center}
\end{figure*}

\begin{figure*}
\begin{center}
\includegraphics[width=0.7\hsize,angle=0]{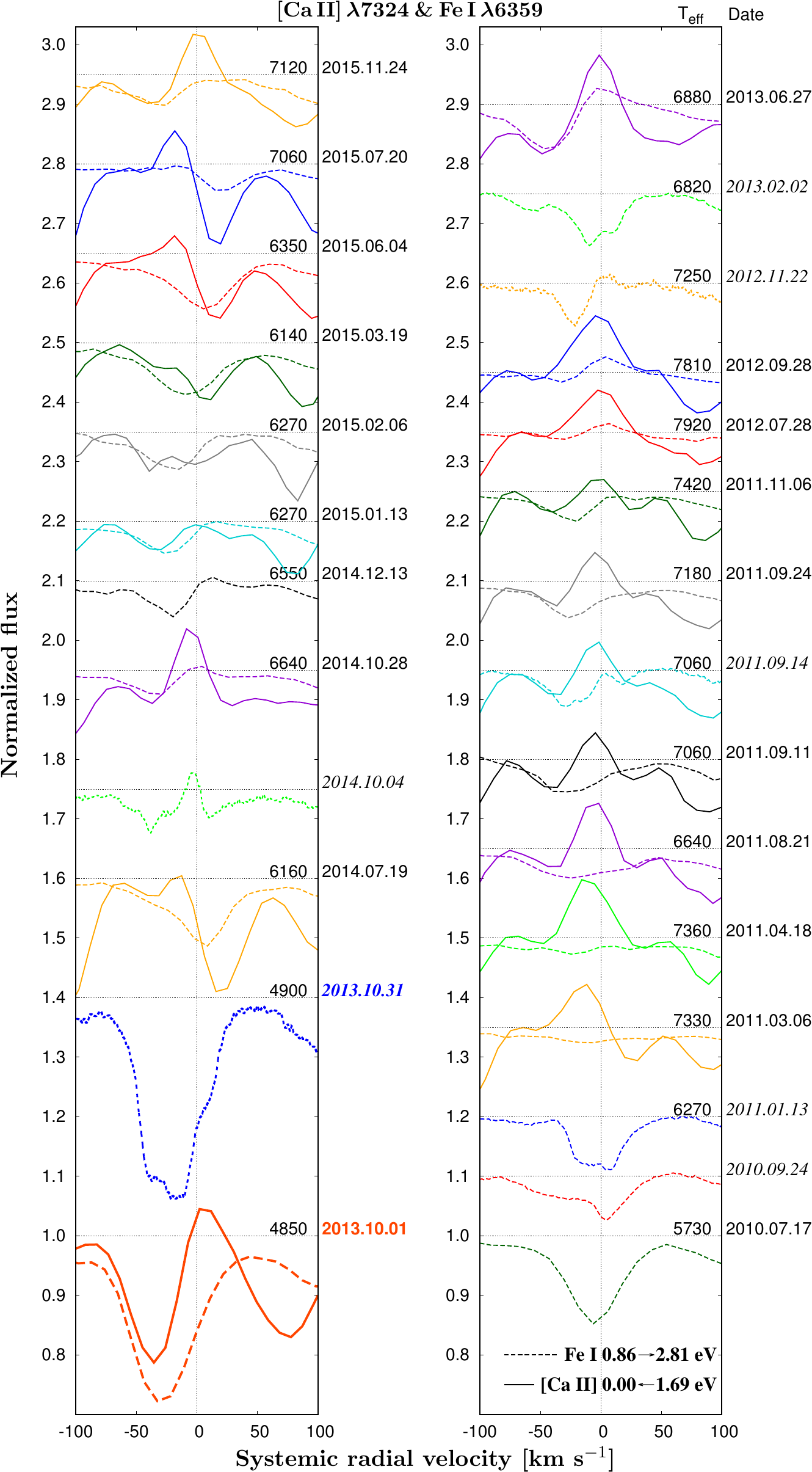}
\caption{Variability in position and intensity of the [Ca\,{\sc ii}] $\lambda$7324 emission line (solid 
lines) and the Fe\,{\sc i} $\lambda$6359 emission line (dashed lines) indicating their synchronous 
behavior.}
\label{fig:app5}
\end{center}
\end{figure*}

In this section we highlight the line-profile variability for selected lines. The lines are 
grouped such that Figures \ref{fig:app1},  \ref{fig:app2}, and \ref{fig:app3} display lines of low,
intermediate, and high excitation potential, respectively. Figure \ref{fig:app4} presents the 
variability in H$\alpha$, and Figure \ref{fig:app5} compares the persistent 
emission in [Ca\,{\sc ii}] $\lambda$7324 with the occasional emission in Fe\,{\sc i} $\lambda$6359.

\bsp	
\label{lastpage}
\end{document}